\documentclass[preprint,review,3p]{elsarticle}

\usepackage{amsmath,amssymb,amsfonts}
\usepackage{algorithmic}
\usepackage{graphicx}
\usepackage{textcomp}
\usepackage{xcolor}
\usepackage{url}
\usepackage{subfigure}
\def\BibTeX{{\rm B\kern-.05em{\sc i\kern-.025em b}\kern-.08em
    T\kern-.1667em\lower.7ex\hbox{E}\kern-.125emX}}

\usepackage{multirow}
\usepackage{tikz}
\usepackage{pgfplots}
\usetikzlibrary{automata, positioning, arrows}
\usepackage{hyperref}

\usepackage[textwidth=3.5cm]{todonotes}

\newtheorem{definition}{\textbf{Definition}}
\newtheorem{proposition}{Proposition}
\newtheorem{assumption}{Assumption}
\newtheorem{remark}{Remark}
\newtheorem{example}{Example}

\newcommand{\TS}{\mathcal{S}}
\newcommand{\StateSet}{\mathcal{X}}
\newcommand{\InitSet}{\StateSet_0}
\newcommand{\TransRel}{\mathcal{E}}
\newcommand{\OutputSet}{\mathcal{Y}}
\newcommand{\OutputMap}{\mathcal{H}}
\newcommand{\TSState}{x}
\newcommand{\TSOutput}{y}
\newcommand{\Post}[1]{\text{Post}(#1)}
\newcommand{\Beh}{\mathcal{B}}
\newcommand{\HBeh}{\Beh^H}

\newcommand{\reals}{\mathbb{R}}
\newcommand{\naturals}{\mathbb{N}}

\newcommand{\CellDomain}{\mathcal{X}}
\newcommand{\MeasDomain}{\mathcal{Z}}

\newcommand{\CurrentDomain}{\mathcal{I}}

\newcommand{\TrainedCont}{\mathcal{K}}

\newcommand{\ClusterRegion}{\mathcal{R}}

\journal{Nonlinear Analysis: Hybrid Systems}

\begin{document}

\begin{frontmatter}

\title{Reinforcement Learning for Robust Ageing-Aware Control of Li-ion Battery Systems with Data-Driven Formal Verification}

\author[label1]{Rudi Coppola\corref{cor1}}
\ead{r.coppola@tudelft.nl}

\author[label1]{Hovsep Touloujian}

\author[label2]{Pierfrancesco Ombrini}

\author[label1]{Manuel Mazo Jr.}
\ead{m.mazo@tudelft.nl}

\cortext[cor1]{corresponding author}

\affiliation[label1]{
organization={Delft University of Technology},
department={Delft Center for Systems and Control}
}
\affiliation[label2]{
organization={Delft University of Technology},
department={Radiation Science and Technology}
}

\begin{abstract}
Rechargeable lithium-ion (Li-ion) batteries are a ubiquitous element of modern technology. In the last decades, the production and design of such batteries and their adjacent embedded charging and safety protocols, denoted by Battery Management Systems (BMS), has taken central stage. A fundamental challenge to be addressed is the trade-off between the speed of charging and the ageing behavior, resulting in the loss of capacity in the battery cell. We rely on a high-fidelity physics-based battery model and propose an approach to data-driven charging and safety protocol design. Following a Counterexample-Guided Inductive Synthesis scheme, we combine Reinforcement Learning (RL) with recent developments in data-driven formal methods to obtain a hybrid control strategy: RL is used to synthesise the individual controllers, and a data-driven abstraction guides their partitioning into a switched structure, depending on the initial output measurements of the battery. The resulting discrete selection among RL-based controllers, coupled with the continuous battery dynamics, realises a hybrid system. When a design meets the desired criteria, the abstraction provides probabilistic guarantees on the closed-loop performance of the cell.
\end{abstract}



\end{frontmatter}

\section{Introduction}

The development of advanced battery control strategies is a critical enabler for the widespread electrification of both energy systems and transportation. These strategies must ensure the safe and reliable operation of batteries under a wide range of conditions throughout their service life. At the same time, growing economic pressures and sustainability goals demand more efficient use of battery systems—minimizing degradation while maximizing performance \cite{weng_fast-charging_2025}. Whether in grid-scale storage or electric vehicles, improving how batteries are charged and managed is central to achieving these objectives.
To this end, significant progress in Battery Management Systems (BMS) is required. Conventional BMS implementations rely heavily on Equivalent Circuit Models (ECMs), consisting of circuital models describing the battery as a set of simplified electrical components. Although ECMs offer fast computation and are widely used in practice, they fail to capture many internal electrochemical processes of the battery. As a result, they are limited in their ability to estimate internal states or inform control strategies that account for degradation mechanisms.

In contrast, physics-based electrochemical models, typically framed as Doyle-Fuller-Newman (DFN) models, offer a more detailed description of the internal dynamics governing battery behavior. These models can, in principle, enable estimation and control of internal variables that are otherwise unobservable, potentially leading to smarter, degradation-aware charging protocols. However, the high computational complexity of DFN models hinders its application in online optimization-based approaches, e.g. nonlinear Model Predictive Control (MPC), or to develop a formal feedback control design.
To bridge this gap, we explore a Reinforcement Learning (RL) approach that leverages the accuracy of physics-based models to construct a light-weight controller by interacting offline with a DFN model simulator. We propose an automatic model-free RL framework that learns optimal charging strategies through interaction with a simulated environment governed by the full electrochemical model. While training relies on DFN simulations, the final controller relies solely on measurable quantities, such as voltage and temperature, rendering our scheme easily applicable in practice. In parallel, we apply a modern statistical tool to provide safety and performance guarantees of the closed-loop system.
This approach offers a promising path toward advanced battery control that is both physically grounded and practically implementable.

A lithium-ion (Li-ion) battery consists of a negative electrode (anode), a separator, and a positive electrode (cathode). The electrodes are porous, composed by microscopic particles, and immersed in a ion-conducting liquid electrolyte. 
Charging occurs as the applied external current forces the internal movement of the ions from the cathode to the anode. 
During discharging the process is reversed.
While this mechanism is highly reversible, enabling hundreds to thousands of charge–discharge cycles, the long-term performance of a battery is constrained by irreversible ageing mechanisms that accumulate over time \cite{ControlOrientedModel}. One of the most prominent degradation processes is the growth of the solid–electrolyte interphase (SEI) layer on the surface of the negative electrode. The SEI layer forms as a result of electrolyte decomposition during the initial cycles and continues to grow during subsequent operation due to parasitic side reactions. As the SEI thickens, it increases interfacial resistance and consumes active lithium, effectively reducing the battery's usable capacity and power capability \cite{AgingMechanisms}.
The kinetics of SEI growth have been modeled as an electron-limited process driven by side reactions at the anode–electrolyte interface, with foundational work provided by Darling and Newman \cite{Darling_Newman_1998}. Importantly, SEI growth is strongly correlated with charging rates: faster charging typically leads to increased SEI formation due to elevated reaction rates and higher overpotentials. This creates an inherent trade-off in charging strategies between minimizing charge time and limiting long-term degradation.
\newline
\textbf{Related work:}
Recent studies have begun to address this by developing ageing-aware charging protocols that balance performance and longevity using physics-based models in a control context, as done in \cite{ControlOrientedModel}. Ageing-aware MPC schemes are implemented on the DFN model, such as in \cite{Khalik_Surrogate, Khalik_DFN}, where an approximate model is used to describe the internal battery dynamics. This work is expanded in \cite{ Khalik_2021}, where a nonlinear MPC framework is implemented using nonlinear optimisation methods. However, the control scheme developed is implemented in full-state feedback as no state estimator concept is developed. As a result, the corresponding experimental runs are performed in an open-loop environment which is detrimental to the charging performance and safety. Additionally, the computational difficulty of MPC based on the DFN model is illustrated in \cite{NMPC_Strategies}, where several problem formulation strategies are considered for the DFN model, each resulting in substantially large time required for a single optimization call (45-65 seconds), which renders real-time MPC on such a complex model infeasible in most practical applications.

Reinforcement Learning (RL) is a model-free approach to policy design. Such a policy, or controller, is designed in \cite{RL_Batteries}, where the DFN Model is used to emulate a cylindrical cell Li-ion Battery System, and a controller is trained for the case of both state feedback and output feedback, i.e., the controller is assumed to only know the cell voltage, the average temperature of the cell, and its State of Charge (SOC). The resulting protocol is shown to perform better than standard rule-based charging methods. The authors model ageing as lithium plating \cite{gao_interplay_2021, lu_multiscale_2023}; accordingly, the cost function does not aim at extending the life cycle of the battery, rather it aims at fast charging while avoiding destructive phenomena. In our work, we model ageing as the SEI growth causing the decay of the State of Health (SOH) of the cell, a measure of its capacity; we explicitly incorporate in the cost function of the RL problem a metric for capacity loss.
In \cite{chowdhury2025adaptive} the authors apply RL to the problem of safe fast charging and augment it with a safety layer that projects unsafe actions into a feasible region defined by a Gaussian Process (GP) surrogate model. In contrast with our work, the authors do not model ageing constraints. Moreover, GP-based approaches rely on the samples used to construct the surrogate model to be Gaussian-distributed, which is often difficult to verify. 
\newline
\textbf{Main contributions:} Different from the works cited above, we focus on providing distribution-free probabilistic performance guarantees, while iteratively constructing and refining (a set of) controllers. We adopt recent advances in data-driven formal methods \cite{coppola2022data,coppola2023data, coppola2024data}, which allow for the construction of a finite abstraction of the battery that conservatively approximates the behavior of the battery under different initial conditions, and manufacturing parameters: this makes the abstraction suitable to describe the properties that the battery satisfies. We demonstrate our approach by synthesizing a safe controller for a time-bounded Reach-While-Avoid specification. In particular, we follow the architecture of Counterexample Guided Inductive Synthesis (CEGIS) \cite{abate2018counterexample, Counterexample_Lyapunov, ahmed2020automated}, where the synthesis of a suitable controller is reached by means of an iterative interaction between a learner and a verifier, detailed in Section \ref{sec:preliminaries}.
From a practical standpoint, we develop controllers trained on a high-fidelity DFN model of the battery cell.  Our framework accounts for the battery's manufacturing parameter uncertainty and State of Health decay, rendering the obtained controller applicable to batteries throughout their entire life span and robust to parameter variation. We obtain a charging protocol with formal guarantees on the maximum rate of ageing of the battery, while ensuring fast and safe charging. The charging protocol is implemented in output feedback relying on realistic output measurements. We compare our solution against the standard Constant-Current-Constant-Voltage (CC-CV) approach as well as against an approach based solely on RL, and we illustrate an improved tradeoff between fast charging and ageing.

\section{Preliminaries}\label{sec:preliminaries}

\subsection{Models and abstractions}
We consider control systems of the form:
\begin{equation}\label{eq::ControlSystem}
    \Sigma_{\text{\text{c}}}=\begin{cases}x_{k+1} = 
        f_{\text{\text{c}}}(x_k,u_k), \\
        z_k = g(x_k), \\
        y_k = \phi(z_k), \\ 
        x_0 = x,
    \end{cases}
\end{equation}
where $x_k\in\CellDomain\subseteq\reals^n$ is the $n$-dimensional state of the cell at time $k\in\naturals$ contained in a domain of interest $\CellDomain$, $g : \CellDomain^n\rightarrow\MeasDomain$ is the measurement map with $\MeasDomain\subseteq\reals^m$ and $m < n$, $z_k$ is the (real-valued) \emph{output measurement}, and $y_k\in \mathcal{Y}$ is the \emph{output label} with $\text{card}(\mathcal{Y})<\infty$, where $\text{card}(\cdot)$ denotes for the cardinality of a set. We employ output measurements for controller synthesis and output labels for verification. The mapping $\phi:\MeasDomain\rightarrow\mathcal{Y}$ is a partitioning map that returns a label, i.e., an element of the finite set $\mathcal{Y}$, corresponding to the output measurement $z_k$.
We seek to synthesize output feedback controllers $\TrainedCont$, i.e., $u_k=\TrainedCont(g(x_k))$, which when applied to \eqref{eq::ControlSystem} results in autonomous systems of the form:
\begin{align} \label{eq::AutonomousSystem}
    \Sigma\doteq \begin{cases}
        x_{k+1} = f(x_k) \doteq f_{\text{\text{c}}}(x_k,\TrainedCont(g(x_k))) 
        \\ 
        y_k = h(x_k)\doteq \phi (g(x_k)),
        \\ 
        x_0 = x,
    \end{cases}
\end{align}
where we have defined $h \doteq \phi\circ g$.
\begin{remark}[Parameter Uncertainty]\label{rem:parameter-uncertainty}
First-principle models, as the ones we employ to model batteries later on, often depend on a number of parameters, that while constant may be difficult to measure and uncertain. Let $p\in \Delta_p$, be a vector of uncertain parameters in a known parametric uncertainty set. Assuming the uncertainties are static, by augmenting the state vector $x$ with $p$, the uncertain system can be reformulated in the same form of \eqref{eq::AutonomousSystem} as:
\begin{align}
    \overline{\Sigma}\doteq \begin{cases}
        \begin{bmatrix}x_{k+1} \\ p_{k+1} \end{bmatrix} = \begin{bmatrix}f(x_k,p_k) \\ p_k\end{bmatrix},\\ y_k = h(x_k),\\ x_0 = x, p_0 = p,
    \end{cases}
\end{align}    
\end{remark}

In this work, we approximate the \emph{concrete} system \eqref{eq::AutonomousSystem}, representing the battery cell in closed loop with a controller, via \emph{abstractions} in the form of finite-state \emph{Transition Systems} (TS), to make them amenable to computer-based verification algorithms \cite{tabuada2009verification}:
\begin{definition}[Transition System \cite{coppola2023data}]\label{def:TS}
    A transition system $\TS$ is a tuple $(\StateSet,\InitSet,\TransRel,\OutputSet,\OutputMap)$, where
    \begin{itemize}
        \item $\StateSet$ is the (possibly infinite) set of states;
        \item $\InitSet \subseteq \StateSet$ is the set of initial states;
        \item $\TransRel\subseteq \StateSet \times \StateSet$ is the set of edges, or transitions;
        \item $\OutputSet$ is the set of outputs;
        \item $\OutputMap:\StateSet \to \OutputSet$ is the output map.
    \end{itemize}
\end{definition}
For any $\TSState\in\StateSet$ its \emph{successor states} are defined as $\Post{x}\doteq\{\TSState'\in\InitSet : (\TSState,\TSState')\in\TransRel\}$. For $\TSState\in\InitSet$ and $H\in\naturals$ we denote by $\HBeh_{\TSState}(\TS)$ the set of all $H$-long output sequence $\TSOutput_0\TSOutput_1...$ stemming from $\TSState$ such that $\TSState_0=\TSState$, $\TSState_{i+1}\in\Post{\TSState_{i}}$, and $\TSOutput(\TSState_i)=\TSOutput_i$ for $i=1,...,H-1$. Similarly, we denote by $\HBeh(\TS)$ the set of all $H$-long output sequences starting from any $\TSState\in\InitSet$. An element $\mathbf{b}$ in $\HBeh(\TS)$ is called \emph{$H$-long behavior} of $\TS$. For a $H$-long behavior $\mathbf{b}=b_0b_1...b_{H-1}$ we denote by $\text{sub}_\ell(\mathbf{b})$ the set of \emph{$\ell$-long subsequences} $\boldsymbol\sigma=\sigma_0...\sigma_{\ell-1}$ such that there exists $j \leq H - \ell$ such that $\sigma_i=b_{j+i}$ for $i = 0,...,\ell-1$.
A TS is \emph{finite-state} if $\text{card}(\StateSet)<\infty$, \emph{deterministic} if for all $\TSState\in\StateSet$ it holds that $\text{card}(\Post{x})\leq 1$.
It is easy to check that the model described in \eqref{eq::AutonomousSystem} can be equivalently described as an infinite-state deterministic TS, from here on denoted by $\TS$, where the transition relation is defined as $\TransRel\doteq\{(x,x') : x' = f(x)\}$, $\OutputMap \doteq h$.

\begin{figure}[]
     \centering
     \tikzset{
        ->,  
        >=stealth', 
        node distance=2.5cm, 
        every state/.style={thick, minimum width=1.2cm,fill=gray!0}, 
        initial text=$ $, 
        }

\usetikzlibrary{automata, arrows.meta, positioning}

\begin{tikzpicture}[scale=0.75, transform shape, auto, initial text = start]
    
    \small

    \node (q0) [state with output] {
	$y_0y_1$ \nodepart{lower} $y_0$
    };

    \node[state with output, right of = q0] (q1) {${y_1y_1}$ \nodepart{lower} ${y_1}$};

    \node[state with output, below of = q0] (q2) {${y_0y_0}$ \nodepart{lower} ${y_0}$};

    \node[state with output, right of = q1] (q3) {${y_0y_1y_1}$ \nodepart{lower} ${y_0}$};

    \node[state with output, below of = q3] (q4) {${y_0y_0y_1}$ \nodepart{lower} ${y_0}$};
        
    \node[state with output, right of = q3] (q5) {${y_1y_1y_1}$ \nodepart{lower} ${y_1}$};

    \draw (q0) edge[above] (q1);

    \draw   (q1) edge[loop below, above] (q1);

    \draw   (q2) edge[loop right, above] (q2);
    
    \draw[->] (q2) edge[above] (q0);

    \draw[->] (q4) edge[above] (q3);

    \draw[->] (q3) edge[above] (q5);

    \draw   (q5) edge[loop below, above] (q5);



\end{tikzpicture}
     \caption{Illustration of the SA$\ell$CA for Example \ref{ex:salca} for $\ell=2$ (left) and $\ell=3$ (right). The upper portion of every node represents the state, the lower portion represents the state's output according to Proposition \ref{prop:schmuck}. The edges represent the transition relation $\TransRel_\ell$, and every state of the SA$\ell$CA is initial.}
     \label{fig:salca-example}
 \end{figure}
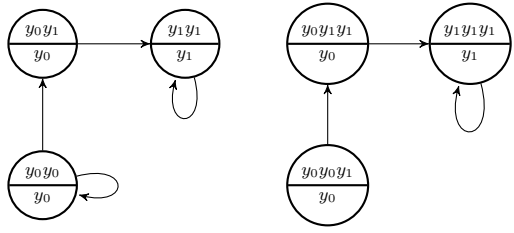

For systems as \eqref{eq::AutonomousSystem}, autonomous with finite output sets, the following result provides a methodology to construct a type of finite-state abstractions, the Strongest Asynchronous $\ell$-complete Abstraction (SA$\ell$CA), capturing in a (possibly conservative) way the external behaviors of the concrete system:
\begin{definition}[SA$\ell$CA~\cite{Schmuck_l_complete}]\label{def:SALCA}
    Let $\StateSet_\ell \doteq \bigcup_{\mathbf{b}\in \HBeh(\TS)}\text{sub}_\ell(\mathbf{b})$ be the set of all $\ell$-long subsequences of $\TS$, and  $\StateSet_{\ell0} \doteq \{\boldsymbol{\sigma}\in\StateSet_\ell : \sigma_0 \in \OutputMap(\StateSet_0)\}$. The TS $\TS_\ell\doteq(\StateSet_\ell,\StateSet_{\ell0},\TransRel_\ell, \OutputSet,\OutputMap_\ell)$ where 
    \begin{align}
        &\TransRel_\ell \doteq \{(\boldsymbol{\sigma},\boldsymbol{\sigma}') \ : \ \sigma_1\sigma_2\ldots\sigma_{\ell-1} = \sigma_0'\sigma_1'\ldots\sigma_{\ell-2}'\}, \label{eq:domino}\\
        &\OutputMap_\ell(\boldsymbol{\sigma}) \doteq \sigma_0. \label{eq:output-ell-map}
    \end{align}
    is defined as the Strongest Asynchronous $\ell$-complete Abstraction (SA$\ell$CA) of system $\TS$.
\end{definition}
The state set of $\TS_\ell$ consists of all the $\ell$-subsequences of $\TS$, and the transitions between states are dictated by the so called \emph{domino rule}: the last $\ell-1$ symbols of a state $\boldsymbol{\sigma}$ must be equal to the first $\ell-1$ symbols of any successor state. The output map is simply given by the first symbol in $\boldsymbol{\sigma}$. While the state set of $\TS$ may be of infinite cardinality, its SA$\ell$CA has necessarily a finite number of states, since there are at most $\OutputSet^{\ell}$ different $\ell$-long subsequences. The main purpose of abstracting an infinite-state system $\TS$ to a finite-state system $\TS_\ell$ lies in the fact that the abstraction allows for a simpler analysis of the behaviors of the original system, as justified by the following proposition.
\begin{proposition}[\cite{Schmuck_l_complete}]\label{prop:schmuck}
    For any $\ell$, $H\geq \ell>1$, the SA$\ell$CA $\TS_\ell$ of $\TS$ satisfies:
    \begin{equation}\label{eq:salca-beh-incl}
        \HBeh(\TS)\subseteq\HBeh(\TS_\ell).
    \end{equation}
\end{proposition}
Proposition \ref{prop:schmuck} ensures that all $H$-long behaviors of $\TS$ are also $H$-long behaviors of $\TS_\ell$. Consequently, if we can verify that \emph{all} the behaviors of the abstraction $\TS_\ell$ satisfy a specification of interest, we can infer that so do the behaviors of the original system.
\begin{remark}
    The parameter $\ell$ determines the coarseness of the abstraction and can be chosen arbitrarily. Given $\ell'>\ell$ it can be shown that $\HBeh(\TS)\subseteq\HBeh(\TS_{\ell'})\subseteq\HBeh(\TS_\ell)$, that is, larger values of $\ell$ will result in tighter over-approximations~\cite{Schmuck_l_complete}.
\end{remark}
\begin{example}\label{ex:salca}
    Consider the TS defined by $\StateSet=\InitSet=[0,1]$, $\TransRel=\{(x,x') : x' = 1/2\cdot x\}$, $\OutputSet=\{y_0,y_1\}$, and $\OutputMap(x) = y_0$ for $x\in(1/4,1]$ and $\OutputMap(x) = y_1$ otherwise, and let $H = 4$. The set $\Beh^4(\TS)$ is given by $y_0y_0y_1y_1$, $y_0y_1y_1y_1$, and $y_1y_1y_1y_1$. The corresponding SA$\ell$CAs for $\ell=2,3$ are given in Figure \ref{fig:salca-example}. Observe that $\Beh^4(\TS_{2})$ contains all the behaviors of $\Beh^4(\TS)$ and additionally contains $y_0y_0y_0y_0$, due to the self loop on state $y_0y_0$, which does not exist in the original system. Approximating $\TS$ by its SA$\ell$CA $\TS_{2}$ allows us to obtain a finite-state TS, which is in general easier to analyze, at the cost of introducing behaviors that might not exist in the original system. Instead, $\ell=3$ gives $\Beh^4(\TS_3) = \Beh^4(\TS)$, which means that we can equivalently study the properties of the behaviors of the finite-state $\TS_3$.
\end{example}

Our goal in this work is to address the design of controllers for a time-bounded Reach-While-Avoid (RWA) specification:
\begin{definition}[(Time-bounded) Reach-While-Avoid Specification]\label{def:rwa-spec}
    Consider the TS $\TS$ with state $x\in \CellDomain$, a set of \emph{unsafe} states denoted by $\CellDomain_U$ and its complement set of \emph{safe} states $\CellDomain_S = \CellDomain\backslash \CellDomain_U$, a set of \emph{initial} states $\CellDomain_0\subseteq \CellDomain_S$, a set of \emph{goal} states $\CellDomain_G\subseteq \CellDomain_S$, and a time horizon of interest $H$. The Reach-While-Avoid specification $\psi(\CellDomain_0,\CellDomain_S,\CellDomain_G,H)$ is satisfied if:
    \begin{equation}
        \forall x_0 \in \CellDomain_0, \exists k' \leq H\;\text{such that}\; x_{k'}\in \CellDomain_G \land \forall  k \leq k' \text{ it holds that }x_k\in \CellDomain_S.
    \end{equation}
\end{definition}
RWA specifications demand that any trace starting from the set $\CellDomain_0$ reaches the target set $\CellDomain_G$ in at most $H$ steps, while remaining at all times in $\CellDomain_S$.
From here on, we assume that the following holds $\psi(\InitSet,\StateSet_S,\StateSet_G,H)$:
\begin{assumption}\label{as:alignment}
    $\InitSet$, $\StateSet_S$, and $\StateSet_G$ are \emph{aligned} with the partitioning map $\OutputMap\doteq h$, i.e. there exist $\OutputSet_{\InitSet},\OutputSet_S, \OutputSet_{G}\subset\OutputSet$ such that $\OutputMap(\TSState)\in \OutputSet_{\InitSet} \iff \TSState\in\InitSet$, $\OutputMap(\TSState)\in \OutputSet_S \iff \StateSet_S$, and $\OutputMap(\TSState)\in \OutputSet_{G} \iff \TSState\in \StateSet_G$.
\end{assumption}  
Consider the sets $\StateSet_{\ell S}\doteq \{\boldsymbol{\sigma}\in\StateSet_\ell : \OutputMap_{\ell}(\boldsymbol{\sigma})\in\OutputSet_S\}$, $\StateSet_{\ell G}\doteq \{\boldsymbol{\sigma}\in\StateSet_\ell : \OutputMap_{\ell}(\boldsymbol{\sigma})\in\OutputSet_G$. From Assumption \ref{as:alignment} and \eqref{eq:salca-beh-incl} we have that $\TS_\ell \models \psi(\StateSet_{\ell0},\StateSet_{\ell S},\StateSet_{\ell G},H)$ implies $\TS \models \psi(\InitSet,S,G,H)$.
In other words, if we know exactly the set $\HBeh(\TS)$ we can construct $\TS_\ell$, check if $\TS_\ell$ satisfies the desired property, and by Proposition \ref{prop:schmuck} conclude that $\TS$ satisfies the desired property as well \footnote{In general, the same reasoning applies for temporal logics expressed in the universal fragment of time-bounded Computation Tree Logic.}.

\subsection{Data-driven abstractions}\label{sec:data-driven-salca} 
Usually, one does not have access to $\HBeh(\TS)$. Instead, one may replace (and relax) such an assumption by the following:
\begin{assumption}
The system \eqref{eq::AutonomousSystem} is initialized with initial condition $x_0$ drawn from a probability distribution $\mathbb{P}$.
\end{assumption}
Then, if only a sub-collection of $\HBeh(\TS)$ can be obtained by sampling a \emph{finite number} of i.i.d. $H$-long behaviors of $\TS$, one can leverage a relaxed version of Proposition \ref{prop:schmuck} defining a so-called \emph{data-driven SA$\ell$CA}
\begin{definition}[\emph{Data-driven SA$\ell$CA}]
\label{def:DD-SALCA}
    Consider a TS $\TS=(\StateSet,\InitSet,\TransRel,\OutputSet,\OutputMap)$. Let $(\InitSet,\mathcal{F},\mathbb{P})$ be a probability space from which a set of $N$ $\mathbb{P}$-distributed samples $\TSState^{(i)}$, $i=1,...,N$, are drawn with their corresponding $H$-long behaviors $\widehat{\HBeh(\TS)}\doteq\bigcup_{i\leq N}\HBeh_{x^{(i)}}(\TS)$. 
    Then the \emph{data-driven SA$\ell$CA} is defined with $\hat{\StateSet}_\ell \doteq \bigcup_{\mathbf{b}\in \widehat{\HBeh(\TS)}}\text{sub}_\ell(\mathbf{b})$ as: $$\hat{\TS}_\ell\doteq(\hat{\StateSet}_\ell,\hat{\StateSet}_{\ell0},\hat{\TransRel}_\ell,\OutputSet,{\hat{\OutputMap}}_\ell)$$ with $\hat{\StateSet}_{\ell0}$ , $\hat{\TransRel}_\ell$, and ${\hat{\OutputMap}}_\ell$ as in Definition \ref{def:SALCA}.
\end{definition}
\begin{proposition}[\cite{coppola2023data}]\label{prop:PAC-inclusion}
     For any $\beta\in(0,1)$ it holds that
    \begin{equation}
        \mathbb{P}^N \Big[ \mathbb{P}[\HBeh_{\TSState^{(N+1)}}(\TS) \in \HBeh(\hat{\TS}_\ell)] \geq 1-\varepsilon(s_N^*,N,\beta) \Big] \geq 1-\beta,
    \end{equation}
    where $\mathbb{P}^N$ is the product probability in $(\InitSet^N,\mathcal{F}^N,\mathbb{P}^N)$, $s^*_{N}$ is a positive integer called the \emph{complexity} and $\varepsilon$ is a function of $\beta$, $N$, and $s^*_N$.
\end{proposition}
The exact expression of $\epsilon$ can be found in \cite{ScenarioEpsilon}. The quantity $s^*_N$ can be understood as follows: recall that to construct the SA$\ell$CA, essentially we need to construct the set of sampled $\ell$-sequences $\hat{\StateSet}_\ell$; $s^*_N$ represents the cardinality of the smallest subset of the sampled $H$-long behaviors $\widehat{\HBeh(\TS)}$ such that, when divided in $\ell$-sequences would return the same set $\hat{\StateSet}_\ell$ (for a thorough discussion see \cite{coppola2023data}). Qualitatively, $\epsilon$ increases for larger values of $s^*_N$ and smaller values of $N$ and $\beta$.
The quantity $1-\beta$ is usually called the \emph{confidence}. Note that Proposition \ref{prop:PAC-inclusion} is a distribution-free result, that is, it holds for any $\mathbb{P}$, even if it is unknown, as long as one can draw i.i.d. samples from $\mathbb{P}$.

An intuitive way to interpret Proposition \ref{prop:PAC-inclusion} is the following: the inner layer $\mathbb{P}[\HBeh_{\TSState^{(N+1)}}(\TS) \in \HBeh(\hat{\TS}_\ell)] \geq 1-\varepsilon$ asserts that for a given set of $N$ initial conditions, and corresponding $H$-long behaviors, the probability of drawing a new initial condition that would result in a behavior existing in the data-driven SA$\ell$CA is not lower than $1-\varepsilon$. As we might have been given a set of realization of initial conditions that poorly represents $\mathbb{P}$, the outer layer of probability accounts exactly for this via the parameter $\beta$. Typically the confidence $\beta$ can be chosen very small, e.g. $10^{-6}$, essentially providing with a near certain guarantee that $\mathbb{P}[\HBeh_{\TSState^{(N+1)}}(\TS) \in \HBeh(\hat{\TS}_\ell)] \geq 1-\varepsilon$. 
Proposition \ref{prop:PAC-inclusion} is the probabilistic translation of Proposition \ref{prop:schmuck}, where the initial condition of the original system is randomly distributed, and the abstraction is obtained from sampled trajectories. In practice, $\hat{\TS}_\ell$ provides a behavioral conservative description (up to a probability) of the original system.

\subsection{Counterexample Guided Inductive Synthesis}
\begin{figure*}[h]
    \centering
    \includegraphics[width = 0.75\textwidth]{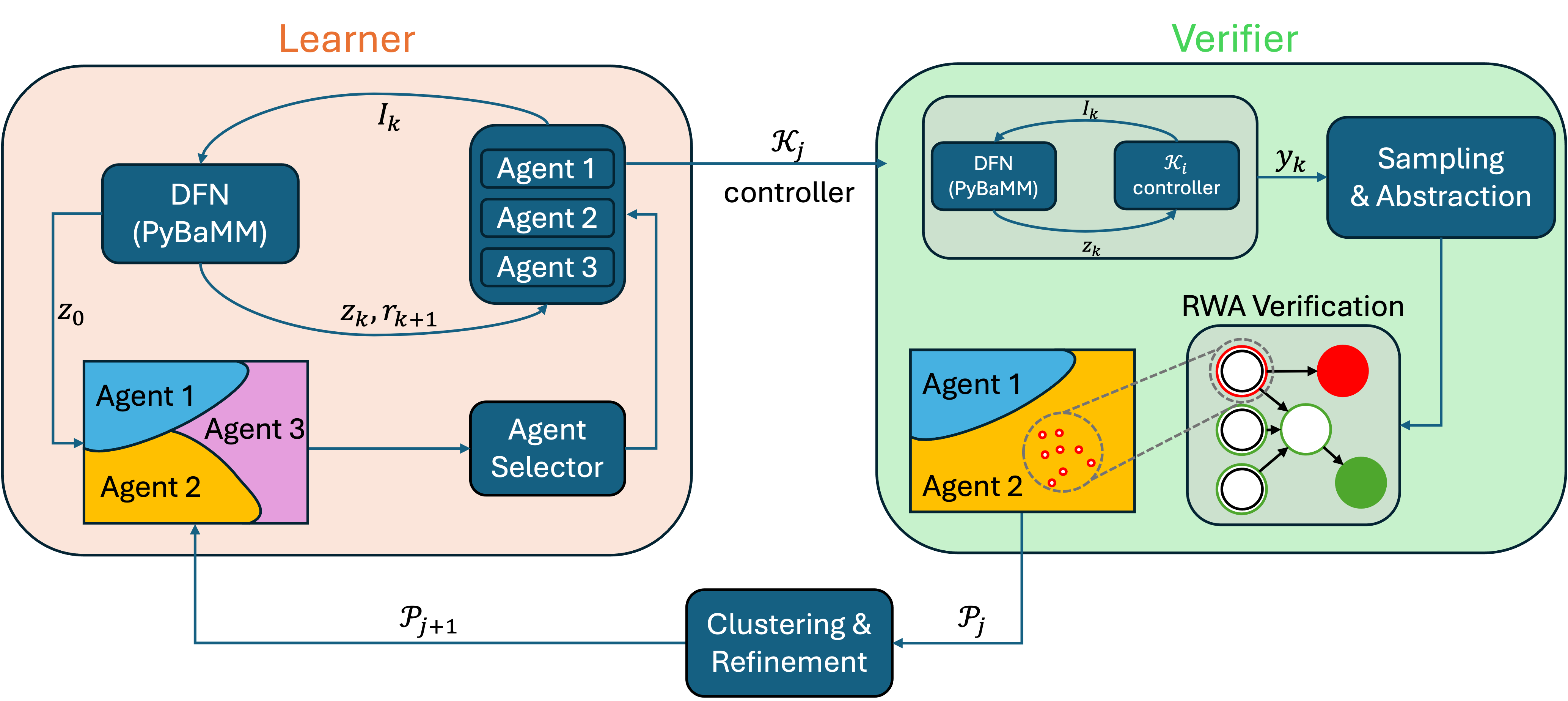}
    \caption{Schematic of the CEGIS architecture. The learner (orange block) proposes a controller $\TrainedCont_j$ based on a number of trained agents, each responsible for a given set of initial conditions. The verifier (green block) uses the closed-loop system with the proposed controller, samples a set of behaviors given by $H$-long output label sequences, and constructs the Data-driven SA$\ell$CA. If the resulting abstraction satisfies the RWA specification the loop terminates with probabilistic guarantees. Otherwise, the domain of initial conditions of the battery cell is partitioned according to the counterexamples.}
    \label{fig:CEGIS_Loop}
\end{figure*}

Counterexample Guided Inductive Synthesis (CEGIS) is an automated procedure which can be used to iteratively construct a controller enforcing a desired specification on a system by learning from counterexamples generated (formally) in the process. CEGIS can be intuitively explained by describing the interaction between its two main blocks, a \emph{learner} and a \emph{verifier}, see Fig.~\ref{fig:CEGIS_Loop}:
\begin{enumerate}
    \item \textbf{Learning (Inductive synthesis) step:} the learner proposes a candidate solution for a specification of interest using any \emph{synthesis tool}. As we detail in Section~\ref{sec:RL}, in our case this step is achieved employing RL to synthesize controllers.
    \item \textbf{Verification step:} the verifier performs a check of the validity of the candidate solution using a \emph{verification tool}. We propose to employ for this step a data-driven SA$\ell$CAs, as presented in Section \ref{sec:data-driven-salca} and further detailed in Section~\ref{sec:data-driven-verification}. 
    If the closed-loop satisfies the RWA specification (with the desired probability and confidence) the synthesis loop is terminated. Otherwise, one must produce counterexamples, in our case initial conditions resulting in $H$-long behaviors violating the specification.
    \item \textbf{Refinement step:} As long as the property is not satisfied, the learner is retrained employing the generated counterexamples and refined solutions (controllers) are proposed. 
\end{enumerate}

\section{Battery Management}\label{sec:physics-modelling}
In this section describe the model of the battery cell's dynamics and the desired specification.
\subsection{Physics-based Cell Modeling}
\begin{figure}[h]
    \centering
    \includegraphics[width = \linewidth]{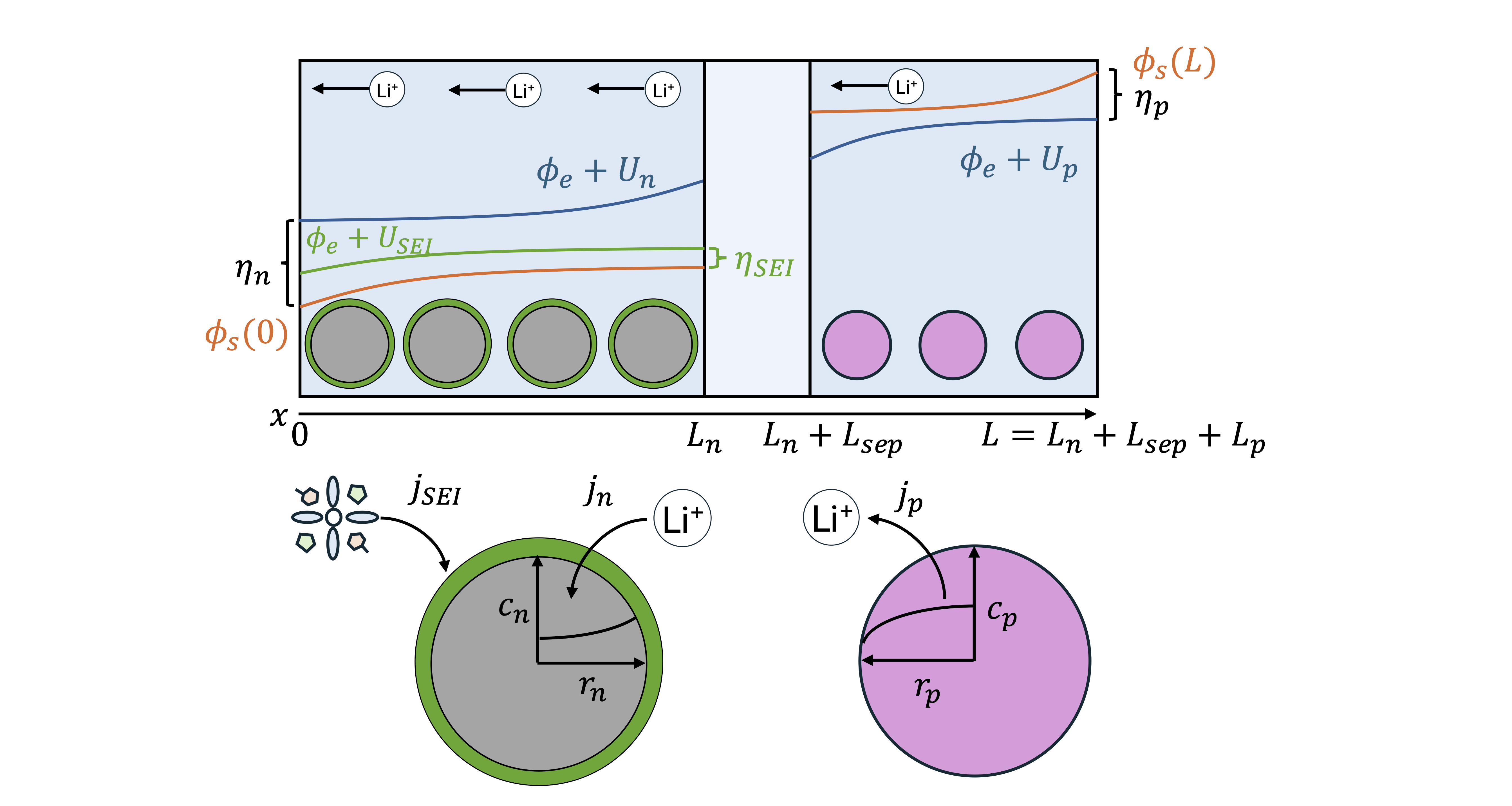}
    \caption{Schematic of the DFN Model. Upper: electrode-scale transport regulates the electrolyte and electric potential drops ($\phi_e$, $\phi_s$), as well as the electrolyte concentration. Consequently affecting the reaction overpotentials ($\eta_n$, $\eta_SEI$, $\eta_p$). Lower: solid diffusion regulates the concentration drop within the particles ($c_n$, $c_p$)}
    \label{fig:DFNSchematic}
\end{figure}

The electrochemical behavior of a Li-ion cell can be described using the physics-based model developed by Doyle, Fuller, and Newman (DFN) \cite{Doyle_Fuller_Newman_1993}. The DFN Model consists of Partial Differential Equations (PDE), describing the spatio-temporal evolution of solid and electrolyte phase concentrations and potentials. Additionally, algebraic constraints are employed to enforce charge and mass conservation. A schematic representation of the DFN Model is shown in Figure \ref{fig:DFNSchematic}.

\subsubsection{Output Dynamics}
The model adopts a \textit{pseudo-two-dimensional} (P2D) approximation based on the DFN model \cite{Doyle_Fuller_Newman_1993, brosa_planella_continuum_2022}. It assumes homogeneous behavior along the lateral and circumferential directions of the cylindrical cell, thereby reducing the problem to a two-scale, one-dimensional model in space. Spatial variation is considered along the through-thickness direction of the cell, denoted by \( x \in [0, L] \), while radial variations in concentration within representative spherical particles are captured using local radial coordinates \( r_n \in [0, R_n]\) and \( r_p \in [0, R_p]\) for the negative and positive electrodes, respectively.

The domain \( [0, L] \) is divided into three contiguous regions: I) negative electrode: \( x \in [0, L_n] \); II) separator: \( x \in [L_n, L_n + L_s] \); III) positive electrode: \( x \in [L_n + L_s, L] \).

Along the electrode thickness, a set of representative spherical particles is placed, used to model lithium diffusion in the solid phase. The decoupling of the \( x \)- and \( r \)-domains is achieved through the interfacial reaction kinetics, which couple the local surface concentration in each particle to the macroscopic current distribution across the cell thickness.

The model evolves over time \( t \) and is governed by the coupled behavior of:
\begin{itemize}
    \item Negative electrode lithium concentration \( c_n(r_n, x, t) \)
    \item Positive electrode lithium concentration \( c_p(r_p, x, t) \)
    \item Electrolyte concentration \( c_e(x, t) \)
    \item Solid-phase potential \( \phi_s(x, t) \)
    \item Electrolyte-phase potential \( \phi_e(x, t) \)
\end{itemize}

The local reactions evolve according to the reaction overpotentials of positive ($\eta_p$) and negative ($\eta_n$) electrodes. These are defined based on $\phi_s$, $\phi_e$ and the surface potentials of the electrodes, $U_n(c_{n, surf})$ and $U_p(c_{p, surf})$. In addition, the model accounts for temperature-dependent properties (e.g., diffusion coefficients, reaction rates, and conductivities), enabling a more realistic description of the electrochemical and thermal dynamics under various operating conditions. The PDEs describing the DFN dynamics can be observed in detail in \cite{chen2020development, Nyman2008}. Hereafter, expressions related to this work are presented.

The cell voltage is defined by the solid-phase potential difference at the two ends of the cell, i.e., 

\begin{equation}
    V(t) = \phi_s(L,t) - \phi_s(0,t).
\end{equation}

The \textbf{State of Charge (SOC)} is defined based on the average lithium concentration in the solid phase of the negative electrode. Defining \( c_{n,\min} \) and \( c_{n,\max} \) as the stoichiometric concentrations corresponding to the voltage cutoffs at 0\% and 100\% SOC, respectively, the SOC is expressed as:

\[
SOC(t) = \frac{\bar{c}_n(t) - c_{n,\min}}{c_{n,\max} - c_{n,\min}},
\]

where \( \bar{c}_n(t) \) is the spatially averaged lithium concentration in the solid particles of the negative electrode at time \( t \).
The temperature variations are assumed to occur homogeneously throughout the cell, resulting in a lumped thermal model. The cell temperature \( T(t) \) evolves according to the energy balance:

\begin{align}
    m c_p \frac{dT}{dt}(t) &= \frac{T_{\text{amb}} - T(t)}{R_{\text{th}}} + q(I_{\text{cell}}, SOC),
\end{align}

where \( m \) is the mass of the cell, \( c_p \) is the effective heat capacity, \( R_{\text{th}} \) is the convective thermal resistance to the environment, and \( q(I_{\text{cell}}, SOC) \) represents the total heat generation rate due to electrochemical processes. In general, the function \( q(I_{\text{cell}}, SOC) \) increases with the magnitude of the applied current and is influenced by the total internal resistance of the cell. However, it also accounts for nonlinear electrochemical effects and thermodynamic heat generation or absorption associated with the electrode reactions. These include ohmic heating, reaction overpotentials, and entropic contributions. A detailed formulation of these heat sources can be found in~\cite{oregan_thermal-electrochemical_2022}.

\subsubsection{Ageing Dynamics}
The model couples ageing dynamics to the lithium intercalation process via a reaction-limited model for the formation of the solid electrolyte interphase (SEI). For the exact mathematical expressions and model parameters, we refer the reader to~\cite{Khalik_DFN, Ramadass2004}; here, we report the qualitative dependence of the relevant variables for simplicity.

The ageing-related \textbf{side reaction current} is denoted by \( j_{\text{SEI}}(x,t) \) and is modeled as

\begin{equation} \label{eq:j2}
    j_{\text{SEI}}(x,t) \propto -j_0(T(t)) \exp\left(-\frac{\eta_{\text{SEI}}(x,t)}{T(t)}\right),
\end{equation}

where $j_0 (T)$ is the empirical exponential pre-factor that depends on temperature ($T$) following an Arrhenius relation, and \( \eta_{\text{SEI}}(x,t) \) is the \textbf{side reaction overpotential} in the negative electrode. It is defined in terms of the electrode potentials and SEI resistance ($R_{\text{SEI}}$) as

\begin{equation} \label{eq:eta2}
    \eta_{\text{SEI}} = \phi_s - \phi_e - U_{\text{SEI}} - (j_n + j_{\text{SEI}}) R_{\text{SEI}}.
\end{equation}

Battery capacity is lost due to the irreversible trapping of lithium ions within the SEI layer, so that the capacity loss \( Q_l \), evolves as 

\begin{equation} \label{eq:cap-loss}
    \frac{dQ_l}{dt} \propto \int_0^{L_n} j_{\text{SEI}}(x,t) \, dx.
\end{equation}

In parallel to the side reaction, the main lithium intercalation reaction in the negative electrode $j_n$ is also governed by an exponential dependence on its own overpotential, defined as

\begin{equation} \label{eq:etan}
    \eta_n = \phi_s - \phi_e - U_n(c_{n,\text{surf}}) - (j_n + j_{\text{SEI}}) R_{\text{SEI}}.
\end{equation}

The evolution of the applied voltage is thus influencing both the intercalation current and the SEI formation dynamics. Mitigating ageing while maintaining fast charge rates requires careful dynamic balancing of the competing intercalation and side reactions, along with effective control of the cell temperature $T$.

\subsubsection{Discretization}
The PDE dynamics describing the cell behavior are discretised in both spatial and temporal domains, and are solved numerically to result in a discrete-time model. Specifically, for the implementation of the model and its numerical solution we rely on PyBaMM, an open-source battery simulation package \cite{sulzer2021python}. According to \ref{eq::ControlSystem}, the discretised representation of the DFN model is given by
\begin{align} \label{eq::OpenLoopSystem}
    \Sigma_{\text{c}}\doteq \begin{cases}
        x_{k+1} = f_{\text{\text{c}}}(x_k,u_k),\\ z_k = g(x_k),\\ x_0 = x,
    \end{cases}
\end{align}
where $x_k\in\CellDomain\subseteq\reals^n$ is the $n$-dimensional state of the cell at time $k\in\naturals$ contained in a domain of interest $\CellDomain$, $u_k\in\CurrentDomain$ is the control input at time $k$ representing the charging current, $g:\CellDomain\rightarrow\MeasDomain$ returns the cell's output measurements. In this work, we assume that $z_k$ comprises the \textbf{time-step, SOC, Voltage, Temperature, Past Input Current}, indicated respectively by $k$, $SOC_k$, $V_k$, $T_k$, $I_{k-1}$, which are typically measurable by the electronics embedded within a BMS. Hereafter, the cell is assumed to be modelled by Equation \eqref{eq::OpenLoopSystem}. Note that we have omitted $y_k$, as it is only used for verification purposes (Section \ref{sec:data-driven-verification}), and it is not used for training the controller.

\subsection{Desired Performance Specification}
With mathematical expressions describing the electrochemical behavior of a Li-ion cell, the control goals of a charging protocol can be formalized in a mathematical framework. 
The control goal is to minimise the loss of capacity $Q_l$ over one charging cycle, while minimising the charging time $t_{final}$.

Concretely, we define a reward function that aims at minimising $Q_l$ and a maximum charging time $t_{\text{max}}$, specifying that the battery needs to reach 90\% SOC with $t_{\text{final}}\leq\ t_{\text{max}}$. 

In addition to the ageing-awareness goal, safety specifications are to be met by maintaining a cell voltage below $V_{\text{max}}$ and a temperature below $T_{\text{max}}$. By maintaining those conditions, the charging protocol can reduce the chances of thermal runaway, or irreversible degradation of the cell components. The above goals can be formalized as a Reach-While-Avoid (RWA) specification.

In the case of a Li-ion Battery cell, the RWA specification can be defined as follows:
\begin{itemize}
    \item \textbf{Initial Set $\CellDomain_0$:} The set of states where $(V, T)\in [\underline{V}, \overline{V}]\times [\underline{T}, \overline{T}]$,
    \item \textbf{Goal Set $\CellDomain_G$:} The set of states where $SOC\in[\underline{SOC}, 1]$,
    \item \textbf{Safe Set $\CellDomain_S$:} The set of states where:
    \begin{enumerate}
        \item Cell Voltage $V \leq V_{\text{max}}$,
        \item Cell Temperature $T \leq T_{\text{max}}$.
    \end{enumerate}
    \item \textbf{Time bound:} The goal must be reached in at most $t_{max}$ steps.
\end{itemize}

The proposed performance must be achieved for a range of batteries having varying manufacturing parameters and state of health (SOH). In order to include cell-to-cell manufacturing variations, the particles' diffusion coefficients, the electrodes' tortuosities and the heat transfer coefficients are varied within a 10\% range following a (clipped) Gaussian distribution. Moreover an SOH between 100 \% and 85 \% is defined as the ratio between the nominal and the aged capacity. The loss of capacity is assumed to be solely due to loss of Li in the SEI layer, so that a corresponding SEI thickness is initialized based on the SOH. Finally the cation transference number is also scaled with the SOH to include ageing-related loss of electrolyte properties. Details are available in \ref{app:params}.

\section{Proposed Approach}\label{sec:proposed-approach}

As introduced in Section~\ref{sec:preliminaries}, we follow the CEGIS approach summarised in Figure \ref{fig:CEGIS_Loop}. In what follows we describe first our \emph{Learning} stage, and our \emph{Verification} (and counterexample generation) stage.

\subsection{Reinforcement Learning}\label{sec:RL}
The learning approach proposed in this work consists of informing the controller design with observation data from a large number of battery simulations (or experiments). We propose to use RL to design a charging protocol in output feedback, similar to \cite{RL_Batteries}. In RL, the goal is to find a policy that maximizes rewards from an environment, in our case the controlled plant: a Li-ion cell. At every time-step $k\in \mathbb{Z}^+$, the environment is described by its state $x_k$ and its output $z_k$, whereas the control policy $\pi(z_k)$ (possibly stochastic) produces an action $u_k$, resulting in the state $x_{k+1}$, following a transition probability $p(x_{k+1}|x_k,u_k)$, and the scalar reward $r_{k+1} = r(x_k,u_k)$. The goal of RL is to learn the optimal policy $\pi^*$ that is associated with the maximum expected reward, i.e., the policy that maximizes the value function
\begin{equation}\label{eq:value-fun}
    V^{\pi}(x) = \mathbb{E}_{\pi}[R_k|x_k=x], 
\end{equation}
where $R_k$ denotes the total reward from time $k$ onward, obtained following a policy $\pi$. In our case the state and action spaces are continuous, motivating the use of the Actor-Critic framework, which is a policy gradient approach that employs function approximators, e.g., neural networks. The parameters describing the actor and critic are updated using the Soft Actor-Critic algorithm \cite{haarnoja2018soft}. This update is done by performing (approximate) gradient descent updates repeatedly, employing simulation (or experimental) data from tuples $(z_k,u_k,r_{k+1},z_{k+1})$, improving the current best policy (actor) and the resulting estimate of \eqref{eq:value-fun}.


The controller operates in Output Feedback, i.e. it measures only SOC, Voltage, Temperature, Past Input Current, and time-step ($SOC_k$, $V_k$, $T_k$, $I_{k-1}$, $k$), while the reward function is designed to balance ageing, charging time, and safety:
\begin{equation}\label{eq:tot-reward}
    r_{k+1} = \lambda_1r_{SOC} + \lambda_2r_{fast} + \lambda_3r_{cap} + r_{final}
\end{equation}
where,
\begin{align}
    r_{SOC} &= SOC_{k+1} - SOC_k, \\
    r_{fast} &= k + 1, \\
    r_{cap} &= -(Q_{\ell,k+1} - Q_{\ell,k}), \\
    r_{final} &=\begin{cases}
        R_{fail} \text{ if unsafe}, \\
        R_{succ} \text{ if goal},\\
        0 \text{ otherwise}.
    \end{cases}
\end{align}
In words, $r_{SOC}$ represents the increase in the SOC between consecutive time-steps, $r_{fast}$ is a penalization of elapsed time, encouraging fast charging, $r_{cap}$ is the capacity loss between consecutive time-steps, as per \eqref{eq:cap-loss}, and $r_{final}$ is the terminal reward of the episode, receiving a large positive value $R_{succ}$ if the SOC safely reaches the goal set $\CellDomain_G$, a large negative one $R_{fail}$ if it violates the safety constraints $\CellDomain_S$, or 0 otherwise. Note that reaching the goal set or exiting the safe set immediately terminates the learning episode, and the environment is reset.

\subsection{Data-driven Synthesis and Verification}\label{sec:data-driven-verification}
After extracting a policy $\TrainedCont_0:\MeasDomain\rightarrow\CurrentDomain$, following the RL scheme in Section \ref{sec:RL}, we formally verify the performance post-training. 
As the reward function \eqref{eq:tot-reward} used to train the RL agent is merely a proxy of the RWA specification of interest, we use a data-driven abstraction to provide safety guarantees.

To this end, we define a finite set of symbols, the output labels, corresponding to different regions of the battery's state; we do this by defining a partition as follows:
\begin{itemize}
    \item The $[0,1]$ interval describing the $SOC$ is finely partitioned in 19 identical sections plus 1 for the Goal set. This is done to avoid self-loops in the abstraction and is pivotal to verifying the reachability of the Goal set, which consists of the region $[\underline{SOC},1]$. The cells or the partition are assigned a unique label, i.e. $a,b,c,...,s,t$, where $t$ labels the Goal set.
    \item The rest of the output variables are partitioned based on their atomic proposition, i.e., the symbol $a$ for safety, such as $V\leq V_{\text{max}}$, $T\leq T_{\text{max}}$. The violation of those constraints results in a symbol $b$.
\end{itemize}
Accordingly, at every time-step $k$ the cell outputs an output label $\TSOutput_k\in\OutputSet$, where $\OutputSet=[a,b,\ldots,t]\times[a,b]\times[a,b]$.

Let $j$ be an integer denoting the current iteration of the CEGIS algorithm, initialised as $j=0$.

\subsubsection*{Step 1 : Data-Driven Abstraction}\label{sec:data-driven-abs}
For a candidate charging protocol $\TrainedCont_i$ we analyze the resulting autonomous system, and, according to equation \eqref{eq::AutonomousSystem}, its dynamics are described by
\begin{equation*}
    \Sigma=\begin{cases}x_{k+1} = 
        f(x)\doteq f_{\text{\text{na}}}(x_k,\TrainedCont_i(g(x_k))), \\
        y_k = h(x_k)\doteq\phi(g(x_k)), \\ 
        x_0 = x^{(i)}.
    \end{cases}
\end{equation*}
Let $D=\bigcup_{i=1}^{N}x_0^{(i)}$ be a set of i.i.d. sampled initial conditions, including random manufacturing parameters and SOHs \footnote{We select the initial voltage and temperature following a uniform distribution on the initial set. In practice, a manufacturer or client may have a data set describing the empirical distribution of the initial conditions: in that case we can easily adapt our sampling to a realistic setting.}, and the resulting $H$-long behaviours $\widehat{\HBeh(\TS)}=\bigcup_{i\leq N}\HBeh_{x^{(i)}}(\TS)$. For a selected $\ell$, construct the TS data-driven SA$\ell$CA $\hat{\TS}_\ell\doteq(\hat{\StateSet}_\ell,\hat{\StateSet}_{\ell0},\hat{\TransRel}_\ell,\OutputSet,\hat{\OutputMap}_\ell)$.

\subsubsection*{Step 2 : Verification of RWA Specification}\label{sec:verif-RWA} Next, we employ the resulting finite-state data-driven SA$\ell$CA $\hat{\TS}_\ell$ to verify the given RWA specification. Solving a Safety Game followed by a Reachability Game~\cite{tabuada2009verification}, one can find the set of counterexamples of initial states in the abstraction $\hat\StateSet_{\ell}^{\text{c}}\in\hat{\StateSet}_{\ell0}$ leading to a violation of the RWA specification. If this set is empty, the desired guarantees are derived according to Proposition \ref{prop:PAC-inclusion}, and the CEGIS iterations are terminated. If $\hat\StateSet_{\ell}^{\text{c}}$ is not empty, we need to extract the associated counterexamples to refine the controller. Note first that the elements $\boldsymbol\sigma \in  \hat\StateSet_{\ell}^{\text{c}}$ are sequences of output labels of length $\ell$. Let us denote by $h^{-1}(\boldsymbol\sigma)$ the set of states that can produce the sequence of symbols $\boldsymbol\sigma=\sigma_0,\ldots,\sigma_{\ell-1}$, i.e. 
\begin{equation}
    h^{-1}(\boldsymbol\sigma):=\lbrace x\in\StateSet\,| \, \boldsymbol{\sigma}\in\Beh^{\ell}_{x}(\TS) \rbrace,
\end{equation} and, abusing notation, $h^{-1}(\hat\StateSet_{\ell}^{\text{c}})=\bigcup_{\boldsymbol{\sigma}\in\hat\StateSet_{\ell}^{\text{c}}} h^{-1}(\boldsymbol\sigma)$.
While computing $h^{-1}(\hat\StateSet_{\ell}^{\text{c}})$ is in general not possible, we can nonetheless extract a set of counterexamples in the domain of the original system $\TS$ from said set using the sampled initial conditions as: \begin{equation}
    \hat\StateSet^{\text{c}}=h^{-1}(\hat\StateSet_{\ell}^{\text{c}})\cap D,
\end{equation}
where $ \hat\StateSet^{\text{c}}\subset\CellDomain$. Increment the counter $j = j + 1$.

\subsubsection*{Step 3: Clustering}
Next, based on the counterexamples  $\hat\StateSet^{\text{c}}$, we generate a partition of the set of initial conditions of the battery cell $\bigcup\{\ClusterRegion^{(m)}\}_{m=1}^{M_j} = \CellDomain_0$ using $M_j > 1$ sets, an arbitrary hyper-parameter of our procedure satisfying $M_j > M_{j-1}$. The increase of clusters per iteration is to ensure that at every new clustering iteration the set of initial conditions $\CellDomain_0$ is divided into increasingly finer partitions. The clustering is performed based on the associated outputs of the counterexamples, i.e. $g(\hat{x}_k)$ for $\hat{x}_k\in \hat\StateSet^{\text{c}}$, to retain the output dependence of the switched controller.
While there exist several algorithms for partitioning \cite{xu2015comprehensive}, in our proposal, for simplicity, the partition is generated by a uniform grid of rectangular hyper-intervals.

\subsubsection*{Step 4: Refinement} Finally, for each region $\ClusterRegion^{(m)}$ we re-train a new agent, for a total of $M_j$ distinct RL agents from which we extract distinct controllers $\TrainedCont_j^{(m)}:\MeasDomain\rightarrow \CurrentDomain$. The resulting controller is defined as
\begin{equation}\label{eq:refined-cont}
    \TrainedCont_j \doteq 
    \begin{cases}
        \TrainedCont_j^{(1)} \text{ if } g(x_0)\in \ClusterRegion^{(1)}, \\
        \ldots \\
        \TrainedCont_j^{(M_j)} \text{ if } g(x_0)\in \ClusterRegion^{(M_j)}.
    \end{cases}
\end{equation}
The resulting $\TrainedCont_j$ can be thought of as a switched controller where the switching depends solely on the measured output at the initial condition, i.e. the region $\ClusterRegion^{(m)}$ containing the initial condition $x_0$ determines the corresponding controller $\TrainedCont_j^{(m)}$ to be applied to the battery cell. Given this construction, return to Step 1.

\section{Results}

We select for demonstration of our approach a battery model based on the LGM50LT ithium-ion cell \cite{chen2020development, oregan_thermal-electrochemical_2022}, which is a commonly used cylindrical cell composed by a Graphite anode and an NMC cathode. The selected battery has a capacity of roughly $5000$mAh and a maximum charging voltage of $4.2$ V. 

We compare our CEGIS resulting protocol with the industry standard Constant-Current-Constant-Voltage (CC-CV) protocol.

\subsection{Benchmarking}

The rule-based Constant-Current-Constant-Voltage (CC-CV) protocol comprises two phases, the Constant-Current (CC) and the Constant-Voltage (CV) phases. During the CC phase a constant charging $I_{\text{CC}}$ is supplied to the cell, until the cell voltage reaches a maximum voltage $V_{\text{CV}}$. Next, the charging protocol transitions to the CV phase, during which the cell's voltage is held constant at $V_{\text{CV}}$, while the current rapidly decays until either the battery is fully charged or a minimum charging current is reached. 

Notably, the CV phase of the protocol is not trivial to maintain, as the current profile that maintains a constant voltage is not directly known and may vary among different batteries. Some systems ensure that the voltage remains at the maximum value by embedding voltage regulators into the circuitry of the Battery Management System, which is prone to high temperature and energy loss.

Given the defined parameter ranges, we identified the cells based on ''quality" and SOH. The quality refers to the manufacturing parameters, with the lowest quality corresponding to the lower bound for the particles' diffusion coefficients and the upper bound for the electrodes' tortuosity. The standard quality corresponds instead to the average parameters of the defined Gaussian distribution. 

\begin{figure}[h]
    \centering
    \includegraphics[width = \linewidth]{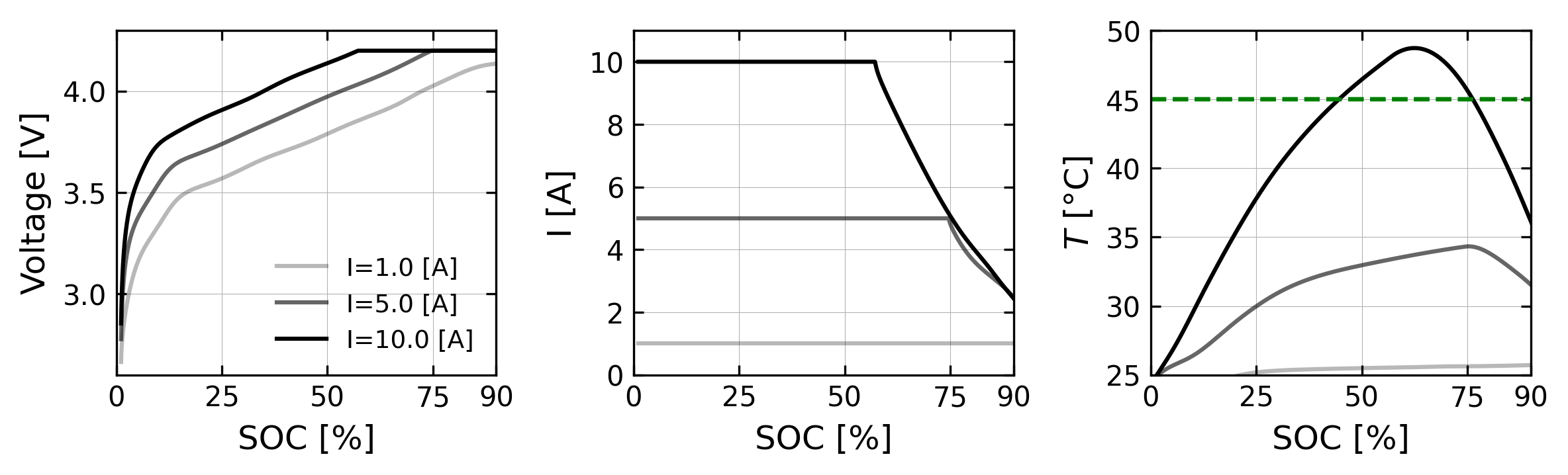}
    \caption{Example of the effects of various input currents on the voltage, current and temperature evolution during a CC-CV protocol.}
    \label{fig:CC-CV_effect_I_traj}
\end{figure}

\begin{figure}[h]
    \centering
    \includegraphics[width = \linewidth]{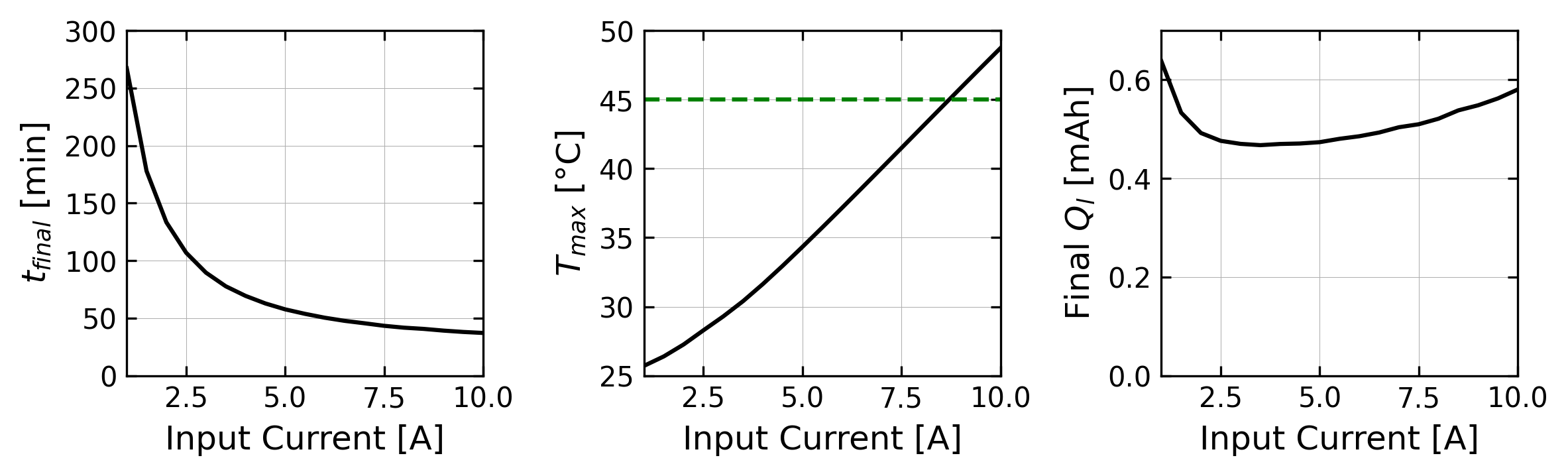}
    \caption{Benchmarking of the CC-CV protocol. Relations between input current and charging times, maximum temperatures and capacity losses. The CC-CV protocol was ended at SOC = 90 \%.}
    \label{fig:CC-CV_effect_I_bench}
\end{figure}

To elucidate the effect of different initial input currents on the performance of the CC-CV protocol, we show in Figure \ref{fig:CC-CV_effect_I_traj} the relation between three different values of current during the CC phase and the relevant variables, for the case of a pristine (SOH = 100 \%) standard quality cell. Figure \ref{fig:CC-CV_effect_I_traj} shows the voltage, current and temperature evolution as a function of the SOC for a CC-CV protocol with the CC phase at $1.0$, $5.0$ and $10.0$ Ampere. The CV phase begins when the voltage of the battery reaches a value of 4.2 Volts. We can observe that the larger the charging current for the CC phase is, the earlier the CV phase is triggered, and the higher the maximum temperature reached. Notably, the temperature trajectory corresponding to a $10.0$ Ampere CC phase violates the safety constraints provided by the battery manufacturer. In Figure \ref{fig:CC-CV_effect_I_bench}, we show the total charging time (from 1\% to 100\%), the maximum temperature reached and the loss of capacity as a function of the input current selected for the CC phase. The charging time decreases with the input current, reaching a lower limit due to the internal resistances within the battery. The maximum temperature reached during the protocol ($T_{max}$) increases roughly linearly with the current input exceeding the safety threshold above 9 A, suggesting the battery should not be operated at this current. The capacity loss is minimal for a current of approximately 3.5 A, coinciding with the recommendation of the producer; accordingly, we select it as the benchmark CC-CV protocol, resulting in $\approx77$ minutes of charging time,  0.47 mAh of capacity loss and a maximum temperature of 30°C.

\begin{figure}[h]
    \centering
    \includegraphics[width = \linewidth]{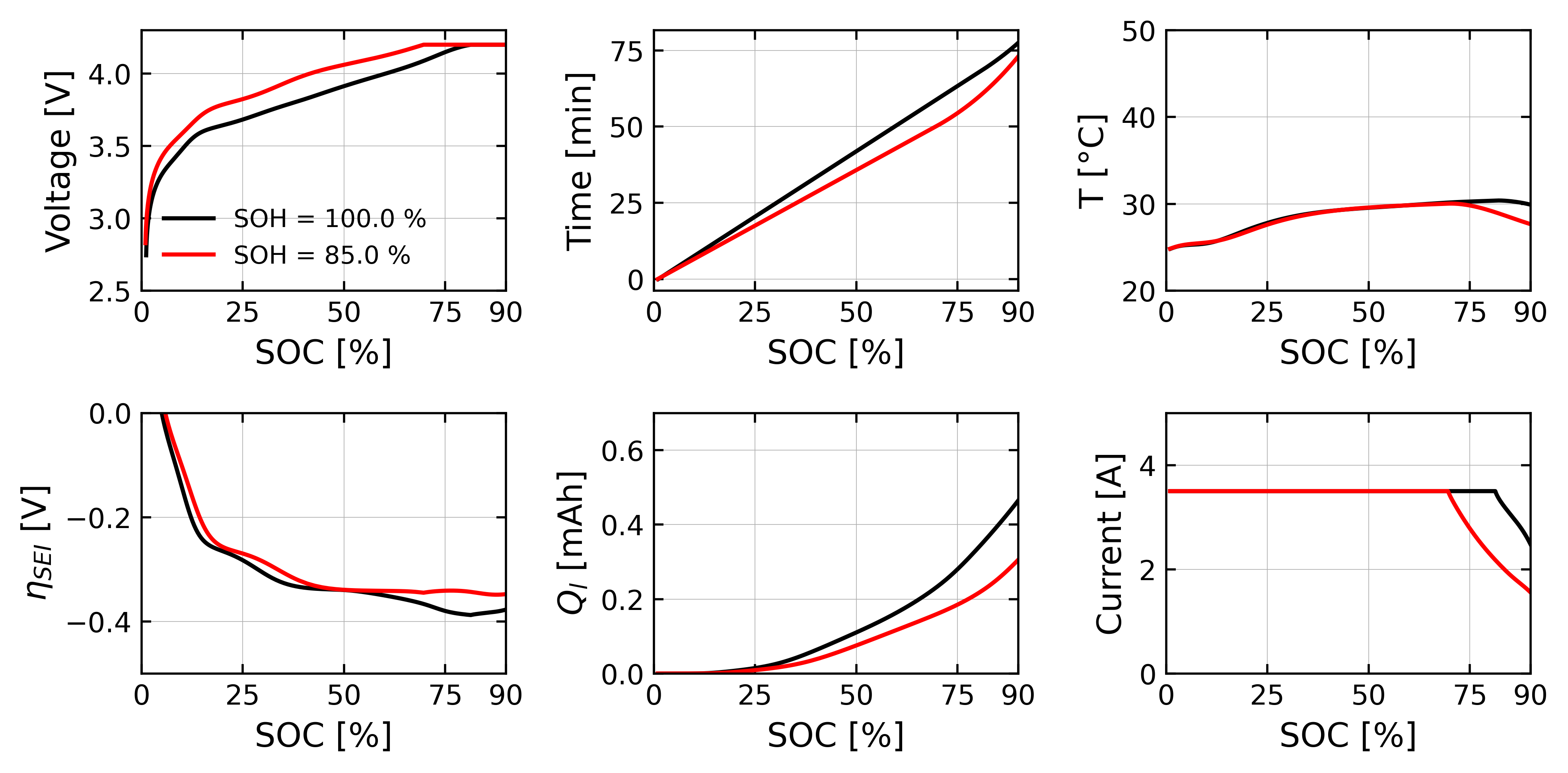}
    \caption{Effect of the CC-CV protocol on pristine and aged cell with average manufacturing parameters.}
    \label{fig:CC_CV}
\end{figure}

To further understand the role of cell SOH, we tested the selected CC-CV protocol (I = 3.5 A) for the case of pristine and aged cells corresponding to an SOH of 100\% and 85\%, respectively. 


Figure \ref{fig:CC_CV} shows the evolution of the variables described in Section \ref{sec:physics-modelling} during the CC-CV protocol. The rate of capacity loss is initially modest, since at low SOCs the negative electrode operates at relatively high voltages. However, once the SOC reaches $\approx15\%$, the low (x-averaged) $\eta_{SEI}$ induces a fast degradation of the electrolyte, accelerating the formation of the SEI layer and the rate of capacity loss. Regulating the negative electrode potential is thus critical to minimise $Q_l$. 


Moreover, applying the same CC-CV protocol to both aged and pristine batteries, we observe marked differences in voltage profiles and degradation kinetics. Notably, while the increased SEI layer thickness leads to higher overpotential, the additional resistance (eq. \ref{eq:eta2}) is also preventing further SEI growth, reducing the additional degradation of the aged battery.

\begin{figure}[h]
    \centering
    \begin{subfigure}
        \centering
        \includegraphics[width=0.475\linewidth]{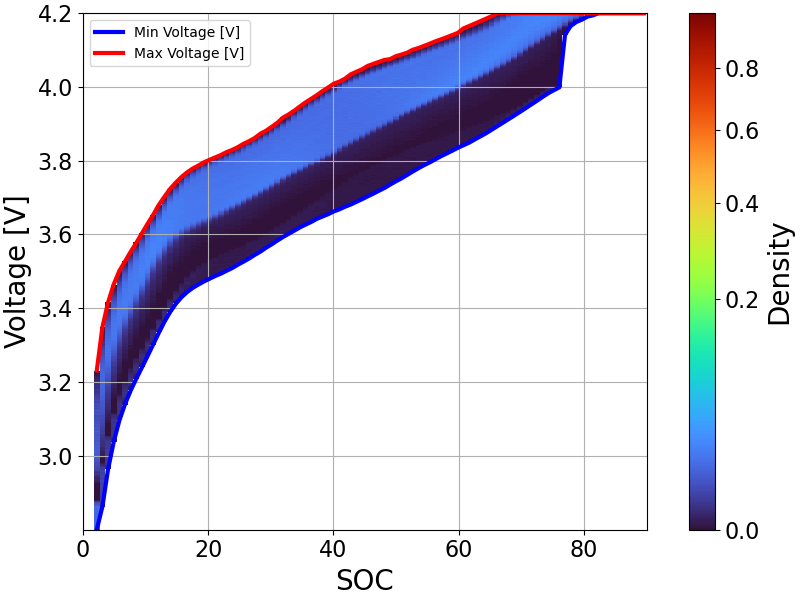}%
        \hfill
        \includegraphics[width=0.475\linewidth]{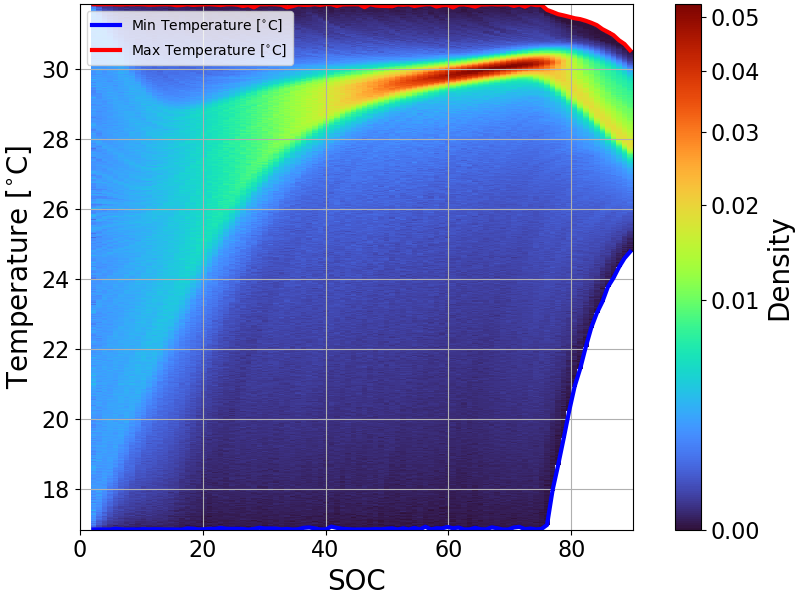}
    \end{subfigure}
    \vskip\baselineskip
    \begin{subfigure}
        \centering
        \includegraphics[width = 0.475\linewidth]{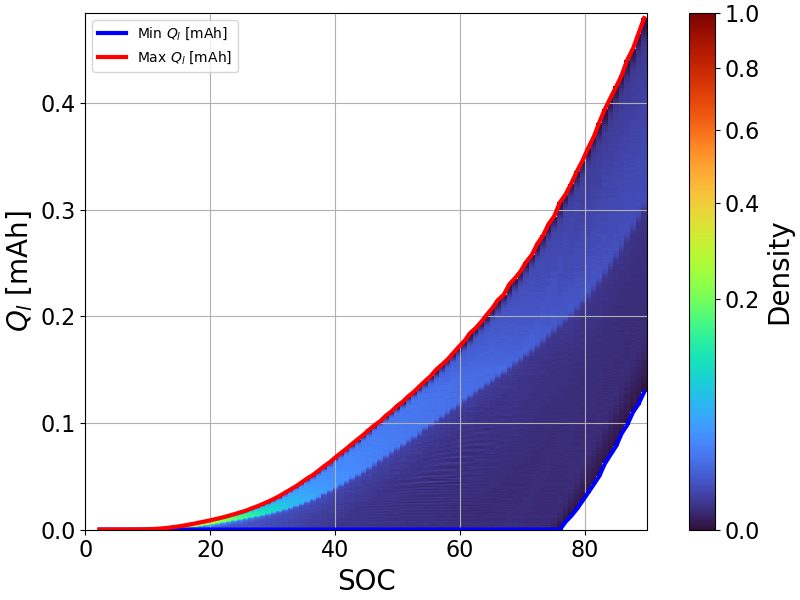}
        \hfill
        \includegraphics[width=0.475\linewidth]{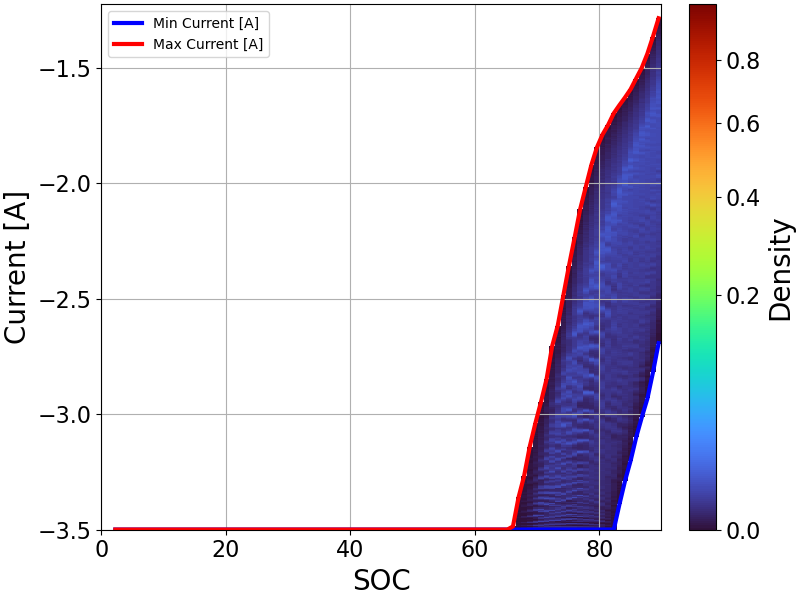}
    \end{subfigure}
    \caption{Trajectory distribution for the CC-CV Protocol}
    \label{fig:CC_CV_Traces}
\end{figure}

In addition, we consider the set of parameters describing a variation on built quality and $SOH \in [0.85, 1.0]$. See \ref{app:params} for the exact parameter ranges. In general, the voltage and temperature profile can vary significantly depending on the SOH and the parameters characterizing the cell's production quality. To illustrate this, we sample $N_{\text{CC-CV}}$ random initial conditions, the parameters characterizing the manufacturing quality and SOH. In Figure \ref{fig:CC_CV_Traces}, we show the trajectories obtained from the sampled conditions and display them as a distribution for different SOCs; specifically, the SOC is binned, and each vertical slice shows the normalised histogram of the measurement within that bin.

Based on the simulations of the CC-CV protocol and the safety constraints provided by the cell manufacturer, we define the RWA specification used for CEGIS:
\begin{itemize}
    \item \textbf{Initial Set $\CellDomain_0$:} The set of states where $(V, T)\in [2.8, 4.0]\times [17, 32]$,
    \item \textbf{Goal Set $\CellDomain_G$:} The set of states where $SOC \in [0.9, 1]$,
    \item \textbf{Safe Set $\CellDomain_S$:} The set of states where:
    \begin{enumerate}
        \item Cell Voltage $V \leq 4.2 V$,
        \item Cell Temperature $T \leq 45^\circ C$.
    \end{enumerate}
    \item \textbf{Time bound:} The goal must be reached in at most 80 minutes of charging time.
\end{itemize}

\subsection{Counterexample-Guided Inductive Synthesis - Results}
Below, we present the results for the scheme described in Section \ref{sec:proposed-approach}. We begin with the first RL run, using a single actor-critic network to control the cell, starting from the all the points in the initial set $\InitSet$.

\subsubsection{Reinforcement Learning Results}
The learning algorithm, along with all subsequent ones, is run using the DelftBlue supercomputer \cite{DHPC2022}. Each training episode is initialised by sampling random initial conditions, manufacturing parameters and SOH.

Referring to Figure \ref{fig:CEGIS_Loop}, the first iteration of our CEGIS scheme starts from the Learner stage, where a single RL agent is trained by interacting with the discretised DFN model described by Equation \eqref{eq::OpenLoopSystem}. The agent selects a new action $I_k$ every 15 seconds of the charging protocol: the DFN model implemented in PyBaMM is integrated using the selected current over the duration and returns the output measurements and reward to the agent. We select a set of parameters for the final agent, resulting in a low-frequency, lightweight controller that can be easily deployed in real-world BMSs. See Table \ref{tab:params} for a summary of the main parameters used for training. When the training is completed, the first candidate controller $\TrainedCont_0$ is ready for testing, and it is parsed to the Verifier stage. We close the loop using the obtained output feedback controller $\TrainedCont_0$, resulting in the autonomous system described by Equation \eqref{eq::AutonomousSystem}. We sample $10^5$ random initial conditions, manufacturing parameters, and SOH, simulate the closed-loop system for a horizon $H=320$ discrete time-steps (corresponding to 80 minutes of maximum charging time), and collect the resulting set of $H$-long behaviors $\widehat{\HBeh(\TS)}$. We select $\ell=6$, construct the data-driven SA$\ell$CA, and check whether it satisfies the RWA specification. We observe that the abstraction does not satisfy the desired properties, and according to Section \ref{sec:verif-RWA}, we extract a set of counterexamples.

For the second iteration, as the counterexamples spread rather uniformly on the set of initial conditions, we define a rectangular partition of the set of initial conditions by dividing the initial voltage interval $[2.8, 4.0]$ into 4 identical segments and the initial temperature interval $[17,32]$ into 2 identical segments: accordingly, we define 8 independent agents, each of which is trained, in parallel, by drawing initial conditions corresponding to the respective cell, using the same parameters shown in Table \ref{tab:params}.
The resulting controller $\TrainedCont_1$ defined by the 8 agents according to Equation \eqref{eq:refined-cont} is used to close the loop, and proceed to the Verifier stage. We construct the data-driven SA$\ell$CA using the parameters in Table \ref{tab:params}, and verify that it satisfies the RWA specification. No counterexamples are found, concluding the CEGIS loop with a successful design.

\begin{remark}
    In our experiments, a single instance of clustering and refinement is sufficient to achieve the desired specifications. If counterexamples were found in one of the refined rectangular cells, it is sufficient to apply the clustering and refinement steps to that specific region, leaving the remaining cells untouched.
\end{remark}

\begin{table}[h]
\centering
\begin{tabular}{llc}
\hline
\textbf{Category} & \textbf{Parameter} & \textbf{Value} \\ \hline
\multirow{6}{*}{Learning (RL agent)} 
& Actor network & 256x256 \\
& Critic network & 256x256 \\
& Control frequency & $6.67\cdot 10^{-2}$ Hz (1/15s) \\
& Training steps & $1.6\cdot 10^{7}$ \\
& Reward weights & $\lambda_1=10^2,\;\lambda_2=10^{5},\;\lambda_3=2.5\cdot 10^{-2}$ \\
& Number of agents (refined) & 8 \\ \hline
\multirow{6}{*}{Verification (SA$\ell$CA)} 
& Number of trajectories $N$ & $10^{5}$ \\
& Memory length $\ell$ & 6 \\
& Horizon $H$ & 320 (80 min) \\
& Confidence $\beta$ & $10^{-6}$ \\
& Scenario complexity $s^*_{N}$ & 13 \\
& Abstraction's state set cardinality & 166 \\ \hline
\end{tabular}
\caption{Summary of training and verification parameters for RL-based controller synthesis.}
\label{tab:params}
\end{table}

\begin{figure}[h]
    \centering
    \begin{subfigure}
        \centering
        \includegraphics[width=0.49\linewidth]{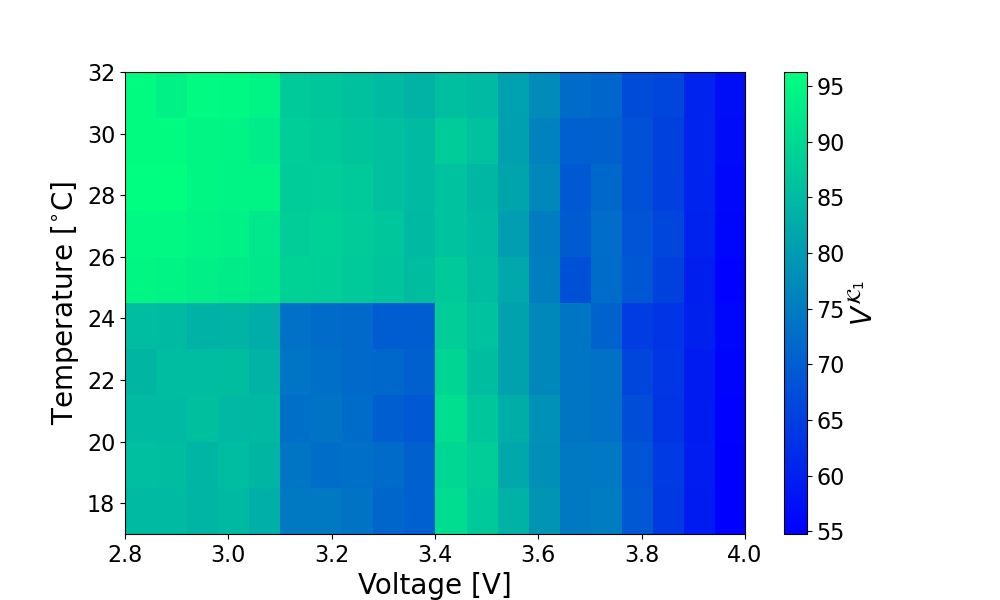}%
        \hfill
    \includegraphics[width=0.49\linewidth]{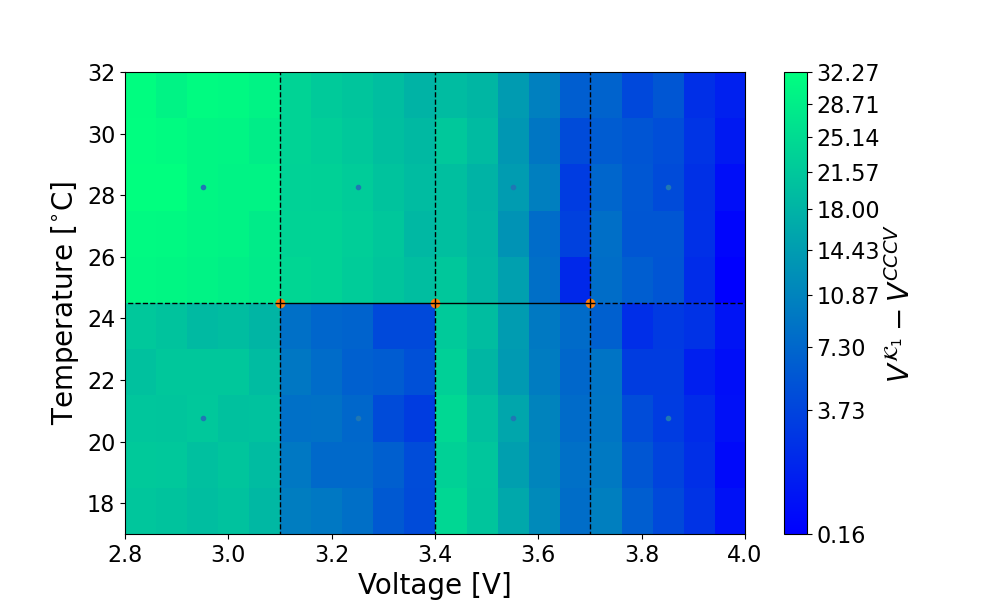}
    \end{subfigure}
    \caption{(Left) Average cumulative reward obtained by the battery controlled with $\TrainedCont_1$. (Right) Average cumulative reward difference between the battery controlled with $\TrainedCont_1$ and a battery controlled by the benchmark CC-CV protocol.}
    \label{fig:RL-avg-reward-performance}
\end{figure}

\begin{figure}[h]
    \centering
    \begin{subfigure}
        \centering
        \includegraphics[width=0.49\linewidth]{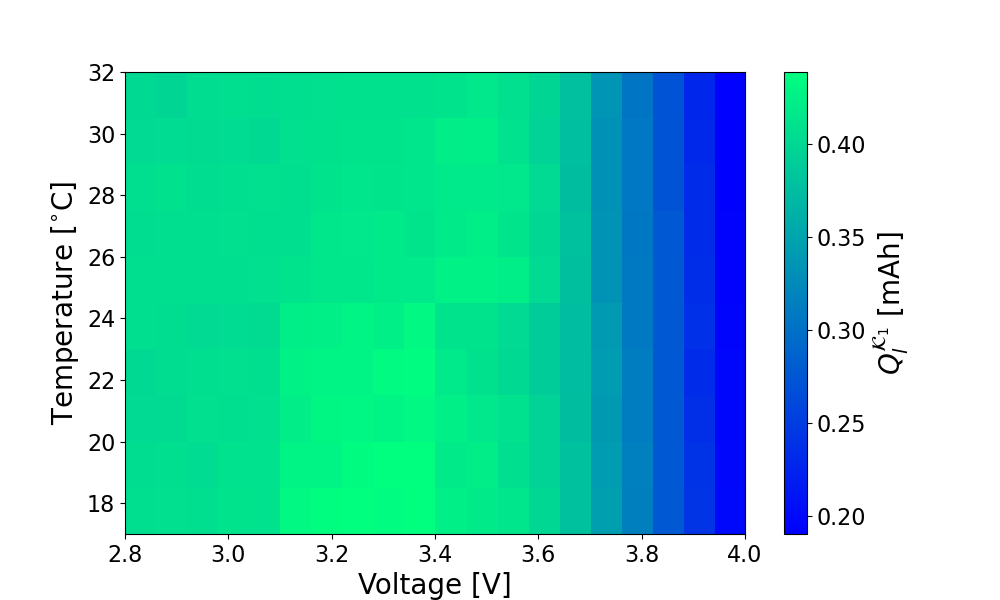}%
        \hfill
    \includegraphics[width=0.49\linewidth]{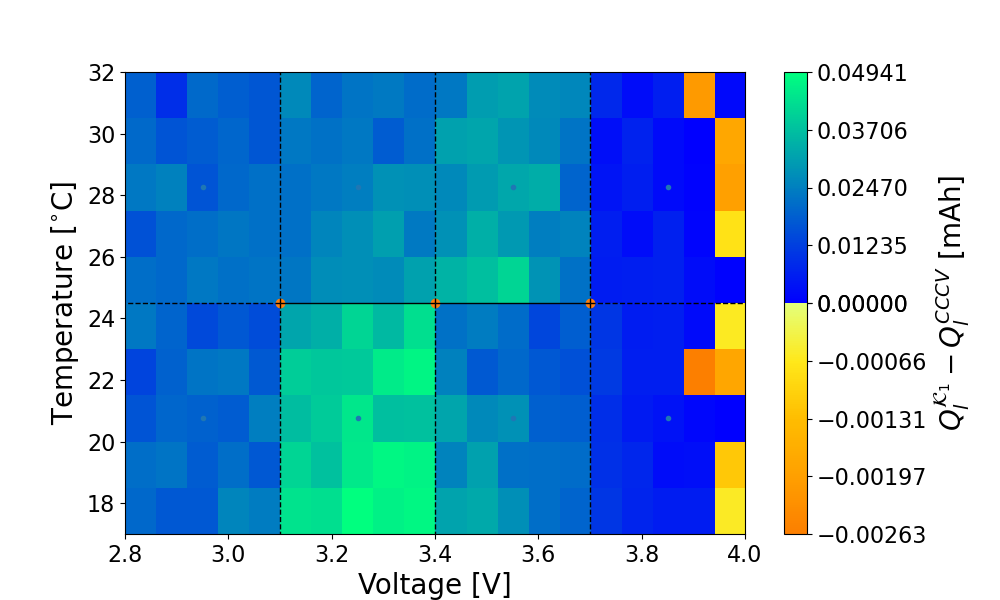}
    \end{subfigure}
    \caption{(Left) Average capacity loss obtained by the battery controlled with $\TrainedCont_1$. (Right) Average capacity loss difference between the battery controlled with $\TrainedCont_1$ and a battery controlled by the benchmark CC-CV protocol.}
    \label{fig:RL-avg-cap-loss}
\end{figure}

\begin{figure}[h]
    \centering
    \begin{subfigure}
        \centering
        \includegraphics[width=0.49\linewidth]{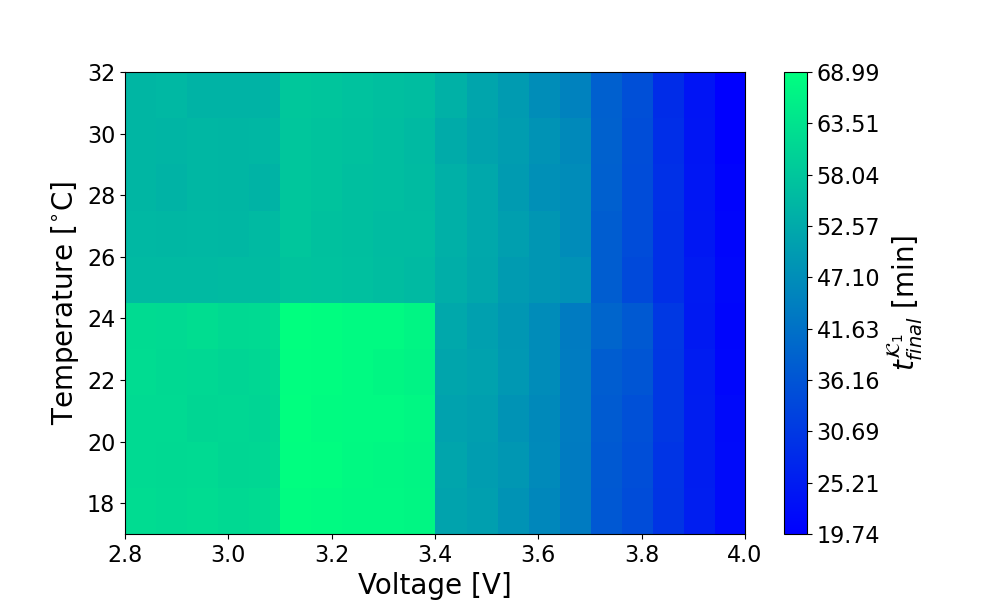}%
        \hfill
    \includegraphics[width=0.49\linewidth]{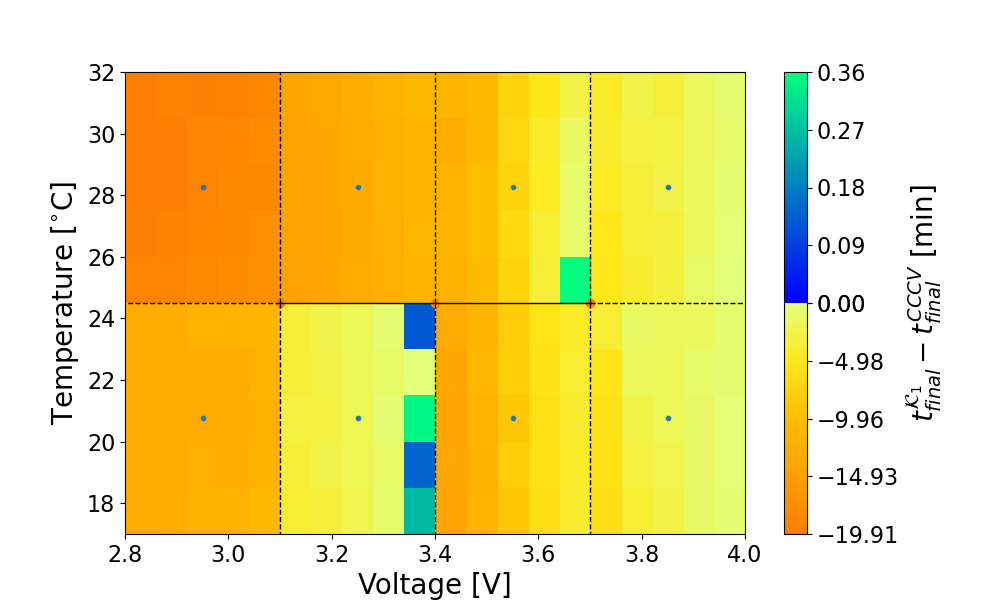}
    \end{subfigure}
    \caption{(Left) Average charging time obtained by the battery controlled with $\TrainedCont_1$. (Right) Average charging time difference between the battery controlled with $\TrainedCont_1$ and a battery controlled by the benchmark CC-CV protocol.}
    \label{fig:RL-avg-chg-time}
\end{figure}

\begin{figure}[h]
    \centering
    \includegraphics[width = \linewidth]{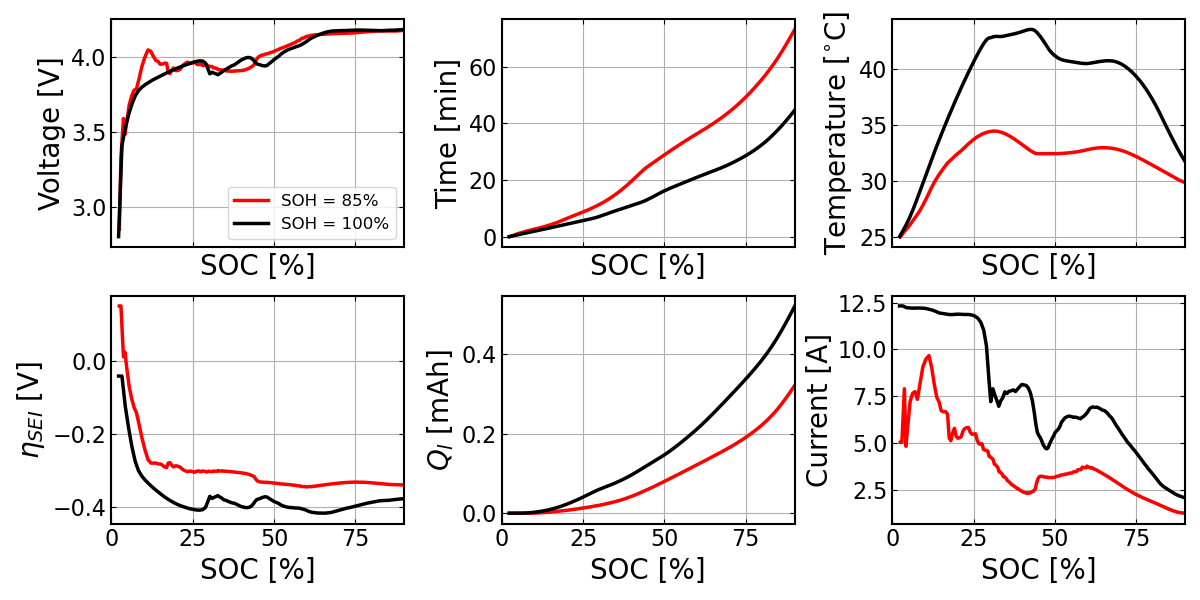}
    \caption{Effect of the CC-CV protocol on pristine and aged cell with average manufacturing parameters.}
    \label{fig:RL_trajs}
\end{figure}

\begin{figure}[h]
    \centering
    \begin{subfigure}
        \centering
        \includegraphics[width=0.475\linewidth]{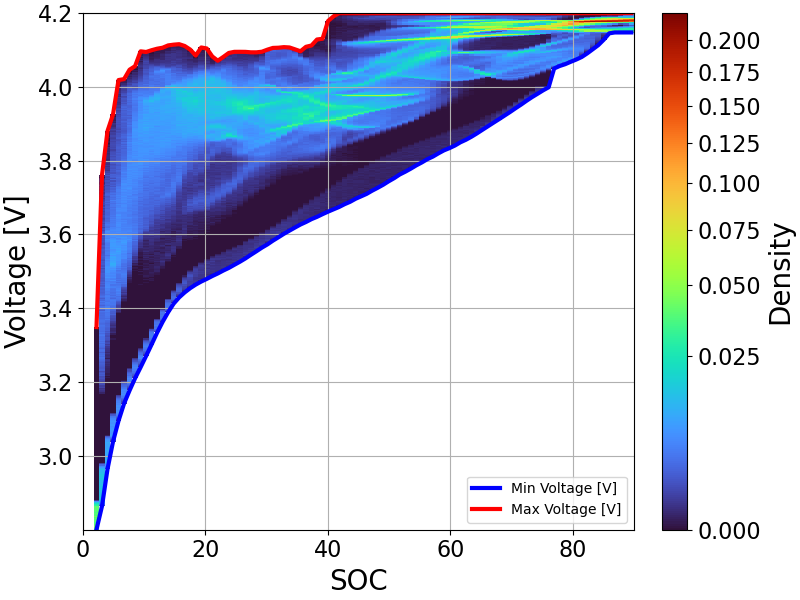}%
        \hfill
        \includegraphics[width=0.475\linewidth]{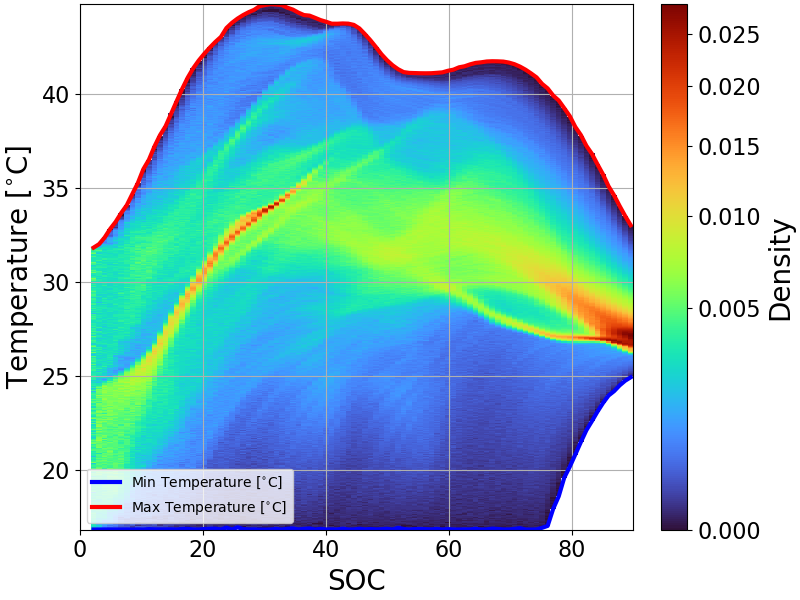}
    \end{subfigure}
    \vskip\baselineskip
    \begin{subfigure}
        \centering
        \includegraphics[width = 0.475\linewidth]{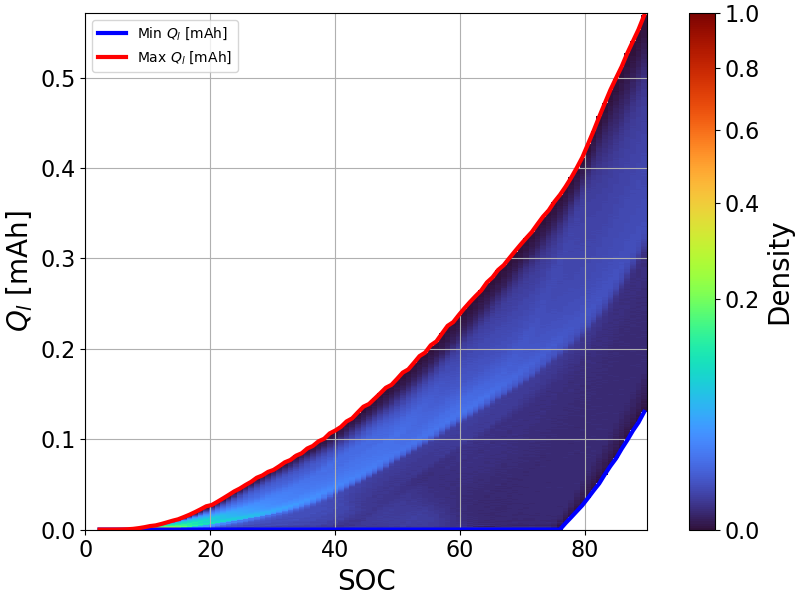}
        \hfill
        \includegraphics[width=0.475\linewidth]{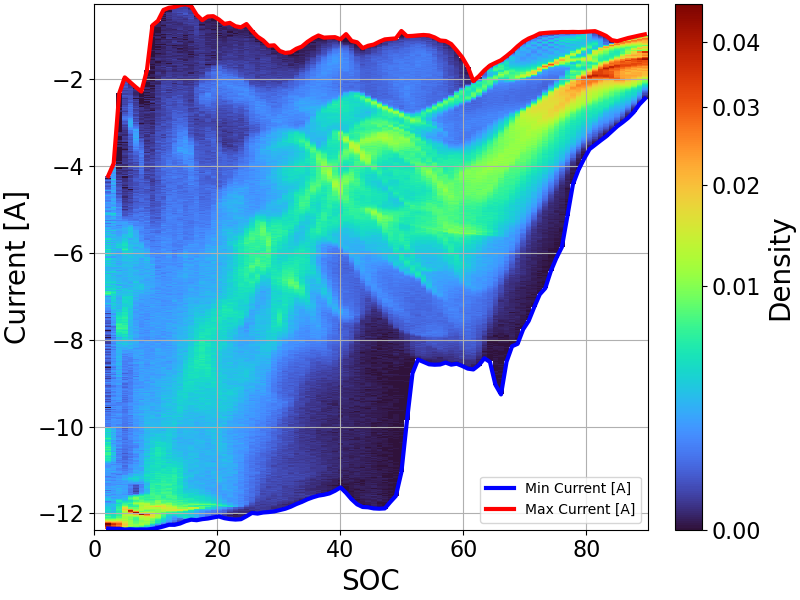}
    \end{subfigure}
    \caption{Trajectory distribution for the final synthesized controller $\TrainedCont_1$.}
    \label{fig:RL_Traces}
\end{figure}

In Figure \ref{fig:RL-avg-reward-performance}, we show the performance calculated as the cumulative reward (representing a trade-off between fast charging and ageing) along all the sampled trajectories: each bin of the heatmap averages the performance across the sampled manufacturing parameters and SOH. We observe that the synthesised controller $\TrainedCont_1$ outperforms the benchmark CC-CV for every initial condition, irrespective of the manufacturing parameters and SOH, when evaluated on the reward function defined in Equation \ref{eq:tot-reward}, where the weights are selected as in Table \ref{tab:params}.   
In Figure \ref{fig:RL-avg-cap-loss}, we further analyse the performance: in particular, we plot the capacity loss measured along charging trajectories as a function of the initial condition in voltage and temperature.
In Figure \ref{fig:RL-avg-chg-time}, we plot the charging time measured along charging trajectories as a function of the initial condition in voltage and temperature. Comparing Figures \ref{fig:RL-avg-cap-loss} and \ref{fig:RL-avg-chg-time} we notice that in the three cells forming the top-left corner, $\TrainedCont_1$ provides essentially the same capacity loss as the benchmark CC-CV protocol, but in a much shorter time, saving up to 20 minutes in total charging time. In the remaining cells, while the reward function is overall better optimised by $\TrainedCont_1$, the incurred cost for the capacity loss is comparable to the incurred saving in charging time.
In Figure \ref{fig:RL_trajs} we plot the evolution of the variables described in Section \ref{sec:physics-modelling} under the charging protocol implemented by $\TrainedCont_1$ with nominal manufacturing parameters and varying SOHs. We observe that, in contrast with the CC-CV trajectories shown in Figure \ref{fig:CC_CV}, $\TrainedCont_1$ applies larger (in magnitude) input currents for lower SOCs, especially on pristine cells. This allows the controller to better utilise the initial section of the charging protocol where $\eta_{SEI}$ is less negative, obtaining a reduction of charging times without additional capacity losses. Accordingly, the temperature rises significantly more, while remaining in the safety constraints. This is also part of the protocol optimisation: higher temperatures reduce the internal resistance of the battery by allowing faster ionic reactions and transport. Notably, the current profile is considerably less smooth, as the reward function does not penalise sudden changes in input current.
Finally, in Figure \ref{fig:RL_Traces}, we show the trajectory distribution for voltage, temperature, capacity loss, and input current as a function of the SOC. In contrast with what we observe in Figure \ref{fig:CC_CV_Traces}, the controller $\TrainedCont_1$ displays more variance across the trajectory distribution, utilising a broader spectrum of current values, particularly for low SOCs. We also observe that $\TrainedCont_1$ spans a larger range of temperature values, always within the safety constraints.
The data-driven SA$\ell$CA constructed in the final step of the CEGIS loop comprises a total of 166 unique $\ell$-sequences, corresponding to the state set of the SA$\ell$CA. The scenario complexity $s^*_{N}$ equals 13, implying that out 
of the $10^5$ collected $H$-long behaviors, 13 of them contain all the 166 unique $\ell$-sequences. With this information, we conclude that  
\begin{equation}
    \mathbb{P}^{N} \Big[ \mathbb{P}[\mathcal{B}_H(\TS(x_0)) \in \mathcal{B}_H(\hat{\TS}_\ell )] \geq 1-4.44\cdot10^{-4} \Big] \geq 1-10^{-6}.\label{eq:final-RL-guarantee}
\end{equation}
In other words, the probability of sampling a new random initial condition (of voltage, temperature, manufacturing parameter and SOH) such that the sequence of output labels forms an $H$-long behavior that does not exist in the set $H$-long behaviors of the data-driven SA$\ell$CA is smaller than $4.44\cdot10^{-4}$, with confidence greater or equal than $99.9999\%$. Since all of the behaviors of the SA$\ell$CA satisfy the RWA specification, we conclude that with probability at least $99.956\%$ and confidence of at least $99.9999\%$ drawing a new initial condition in the battery will return a behavior that also satisfies the RWA specification.

In contrast with established methods for obtaining statistical guarantees using concentration inequalities \cite{reijsbergen2015hypothesis}, generating a full abstraction of the original system has an important advantage: while we have focused on a single specification, the RWA, as we argue in Section \ref{sec:data-driven-salca}, $\hat{\TS}_\ell$ provides a rich behavioral description of the original system $\TS$,  up to the PAC bounds shown above, in the compact form of a finite state abstraction. Consequently, now that we have obtained a successful design, we can reuse $\hat{\TS}_\ell$ to check for other properties of interest; for instance, recalling the partition defined in Section \ref{sec:data-driven-verification}, labels from $g$ to $t$ and from $o$ to $t$ indicate SOCs greater or equal to $50\%$ and $70\%$. From the abstraction, we can verify that every abstract state reaches labels $g$ and $o$ in at most 135 steps and 197 steps, indicating that, with the same probability described in Equation \eqref{eq:final-RL-guarantee}, the battery is guaranteed to reach $50\%$ and $70\%$ of its SOC within at most $33.75$ minutes and $49.25$ minutes respectively, accounting for random initial conditions, manufacturing parameters and SOH.

\section{Conclusion}
In this paper, we presented a novel framework that integrates RL, data-driven formal verification, and CEGIS for the design of ageing-aware charging protocols in Li-ion batteries. By leveraging high-fidelity physics-based models, our approach enables the synthesis of switched output-feedback controllers that explicitly trade off charging speed against long-term degradation while guaranteeing safety with respect to voltage and temperature limits.

A key contribution of this work lies in the use of data-driven abstractions to provide probabilistic, distribution-free guarantees on closed-loop behavior, robust to manufacturing parameter variations and to the dynamics' variations due to ageing. The incorporation of CEGIS further allowed us to iteratively refine candidate controllers in response to counterexamples, ultimately yielding a protocol that significantly improves upon the standard CC–CV method. Our results demonstrate not only faster charging times but also reduced ageing, validated with statistical confidence and robustness to manufacturing variability and state-of-health differences.


While we have relied on a conventional DFN model with SEI-driven capacity fade, our framework is not limited to this specific choice. More advanced electrochemical and ageing models, incorporating, for example, concentration-dependent diffusivities \cite{karanth_phase_2024}, coupled ion–electron transfer kinetics \cite{bazant_unified_2023}, or phase-field representations of phase-separating materials \cite{bai_suppression_2011, ombrini_kinetically_2025, ombrini_thermodynamics_2023, lu_multiscale_2023}, could be employed to more accurately capture cell dynamics across different chemistries and operating regimes \cite{ombrini_kinetically_2025, ombrini_thermodynamics_2023}. From a control perspective, such models would primarily serve as richer training environments, challenging the synthesis pipeline with additional nonlinearities, coupled degradation mechanisms, e.g., lithium plating \cite{gao_interplay_2021, lu_multiscale_2023} or mechanical stress \cite{karger_modeling_2024}, and nontrivial observability issues. Although this would increase the computational cost of training and verification, it would provide an even more stringent validation of the proposed design methodology and ultimately improve its applicability to real-world cells and chemistries.

Future work will focus on extending our framework to such advanced models and validating the resulting controllers experimentally, thereby further bridging the gap between formal, model-driven control design and practical battery management systems.
\appendix

\section{Parameters, cell quality and SOH}\label{app:params}

We use the \texttt{Chen2020} PyBaMM \cite{PyBAMM} parameter set as the nominal model \cite{chen2020development}. Thermal properties (heat capacities, thermal conductivities, densities, and OCP entropic changes) and the Arrhenius activation energies for solid diffusivities are taken from \texttt{O'Regan2022}\cite{oregan_thermal-electrochemical_2022}. Solid diffusivities follow
$
D_{\mathrm{s},k}(T)=D_{\mathrm{s},k}^{\mathrm{ref}}\;d_k\;
\exp\!\left[\frac{E_{D,k}}{R}\Big(\frac{1}{298.15}-\frac{1}{T}\Big)\right],
$
with $k\in\{\mathrm{p},\mathrm{n}\}$ and $E_{D,\mathrm{p}}=12084$ $J mol^{-1} K$, $E_{D,\mathrm{n}}=17447$ $J mol^{-1} K$.

\subsection{Cell quality}
The nominal parameters were modified within a defined range to account for cell-to-cell heterogeneity, using truncated Gaussian distributions.

\begin{table}[h]
\centering
\small
\renewcommand{\arraystretch}{1.15}
\begin{tabular}{l l l}
\hline
\textbf{Quantity} & \textbf{Variation (relative to nominal)} & \textbf{Bounds} \\
\hline
Total heat transfer coefficient & $h_{\text{tot}}/h_{\text{tot},0} \sim \mathcal{N}(1,0.03)$ & $[0.9,\,1.1]$ \\
Solid diffusivity & $D_{\mathrm{s},k}/D_{\mathrm{s},k,0} \sim \mathcal{N}(1,0.03)$ & $[0.9,\,1.1]$ \\
Bruggeman coefficient (electrolyte) & $b_{k}/b_{k,0} \sim \mathcal{N}(1,0.03)$ & $[0.9,\,1.1]$ \\
\hline
\end{tabular}
\caption{Uncertainty factors applied to selected parameters, expressed as ratios to their nominal values. 
Here $\mathcal{N}(1,0.03)$ denotes a normal distribution with mean $1$ and standard deviation $0.03$. 
The subscript $k \in \{\mathrm{p},\mathrm{n}\}$ indicates positive and negative electrodes.}
\end{table}

\subsection{SOH adjustments}
The variation in SOH was accounted for by considering the effect of SEI formation and electrolyte degradation.
We map degradation to transport and capacity as:
\[
t_+ = \text{SOH}\cdot t_{+,0},\qquad
c_{\mathrm{s,n}}(x,0)= \text{SOH}\cdot c_{\mathrm{s,n},0},\qquad
Q_{\mathrm{nom}}=\text{SOH}\cdot Q_{\mathrm{nom},0}.
\]
Uniform SEI thickening links capacity loss to added SEI volume:
\[
\delta_{\mathrm{SEI},0}=\delta_{\mathrm{SEI,init}}+
Q_{\mathrm{nom}}\,(1-\text{SOH})\,
\frac{\bar v_{\mathrm{SEI}}}{a}\,
\frac{3600}{F}\,
\frac{1}{z\,V_{\mathrm{electrode}}},
\]
where $\bar v_{\mathrm{SEI}}$ is the partial molar volume of the SEI, $a=3\varepsilon_{\mathrm{act,n}}/R_{\mathrm{n}}$ and $V_{\mathrm{electrode}}=h\,w\,L_{\mathrm{n}}$, with $h$ and $w$ being the height and width of the electrode.

\bibliographystyle{elsarticle-num} 
\bibliography{my_bib}

\end{document}